\documentclass[a4paper,fleqn]{cas-dc}

\usepackage{lineno}
\usepackage[numbers]{natbib}
\usepackage{xcolor} 
\usepackage{mhchem}
\usepackage{booktabs} 
\usepackage{siunitx}
\def\tsc#1{\csdef{#1}{\textsc{\lowercase{#1}}\xspace}}
\tsc{WGM}
\tsc{QE}

\begin{document}

\let\WriteBookmarks\relax
\def\floatpagepagefraction{1}
\def\textpagefraction{.001}

\shorttitle{Oxidation resistance of 2D GeO$_2$}    

\shortauthors{X. Zhang, X. Yu, L. Ma, Y. Ge, Y. Liu, and W. Wan}  

\title[mode = title]{First principles study on the oxidation resistance of two-dimensional intrinsic and defective GeO${_2}$}  



%


\author[1]{Xixiang Zhang}
\credit{Methodology, Software, Formal analysis, Writing - Original Draft}

\author[1]{Xinmei Yu}
\credit{Formal analysis}

\author[1]{Liang Ma}
\credit{Formal analysis}

\author[1]{Yanfeng Ge}[orcid=0000-0002-2441-7295]
\credit{Writing - review $\&$ editing, Funding acquisition}

\author[1]{Yong Liu}[orcid=0000-0002-5435-9217]
\credit{Writing - review $\&$ editing, Funding acquisition}

\author[1]{Wenhui Wan}[orcid=0000-0002-6824-0495]
\credit{Conceptualization, Funding acquisition, Review \& Editing, Supervision}
\cortext[1]{Corresponding author}
\cormark[1]
\ead{wwh@ysu.edu.cn}

\affiliation[1]{organization={State Key Laboratory of Metastable Materials Science and Technology $\&$ Hebei Key Laboratory of Microstructural Material Physics, School of Science, Yanshan University.},
	addressline={No. 438 West Hebei Avenue}, 
	city={Qinhuangdao},
	postcode={066004}, 
	state={He Bei},
	country={China}}
    



\begin{abstract}
Although two-dimensional (2D) oxide semiconductors exhibit remarkable oxidation resistance compared to conventional 2D materials, the microscopic physical processes that govern this behavior at the atomic scale remains elusive. Using first-principles calculations, we investigated the defect formation and oxidation dynamics of the GeO${_2}$ monolayer (ML). The investigations reveal that the intrinsic GeO${_2}$ ML is resistant to oxidation due to strong electrostatic repulsion between surface oxygen ions and approaching O$_2$ molecules, effectively suppressing chemisorption. In contrast, defective GeO$_2$ ML with surface O vacancies shows vulnerability to oxidation with the O$_2$ molecule occupying the vacancy through a low-energy activation energy ($E_a$) of 0.375 eV. Remarkably, the subsequent O$_2$ dissociation into atomic species faces a higher activation barrier ($E_a$ = 1.604 eV), suggesting self-limiting oxidation behavior. Electronic structure analysis demonstrates that oxidation primarily modifies the valence bands of defective GeO${_2}$ MLs through oxygen incorporation, while the conduction bands and electron effective mass recover to pristine-like characteristics. We further proved that the high O$_2$ pressure hinders the formation of the O vacancy, while high temperature increases the oxidation rate in GeO$_2$ ML. These atomic-level insights not only advance our understanding of oxidation resistance in 2D oxides but also provide guidelines for developing stable GeO${_2}$-based nanoelectronic devices.
\end{abstract}


\begin{highlights}
\item The intrinsic GeO${_2}$ ML has excellent oxidation resistance. 
\item Defective GeO${_2}$ ML with O vacancies becomes vulnerable to O$_2$.
\item The O$_2$ molecule chemisorbs on defective GeO${_2}$ ML without dissociation. 
\item Oxidation restores the effective electron mass of defect GeO${_2}$ ML to the intrinsic one.
\item High temperature speed up the defect formation and oxidation in GeO$_2$ ML.
\end{highlights}


\maketitle

\section{Introduction}
Two-dimensional (2D) semiconductors have great potential for the fabrication of high-performance electronic devices due to their dangling-bond-free surfaces, ease of modulation, tunable electronic properties, and short-channel effects \cite{Chhowalla2016,Sheng2023}. 2D semiconductors offer a feasible solution to scaling limitation in conventional silicon-based technologies, promoting industrial and economic development~\cite {Kim2024}. However, their high surface-to-volume ratio makes them more susceptible to environmental influences than their bulk counterparts \cite{Yang2022,Li2019}. In particular, oxidation can affect the lattice structure, intrinsic physical characteristics, and device performance. For example, 2D InSe \cite{Yang2019} and black phosphorus (BP) \cite{Wu2021} exhibit optimal band gaps and ultrahigh cattier mobility, but their lattice structure degrades soon after oxygen exposure \cite{Wells2018,Ahmed2017}. Their superior semiconducting properties are thus lost in the air. Consequently, enhancing oxidation resistance remains a critical challenge for electronic devices constructed with these 2D semiconductors.

In addition to encapsulation strategies to prevent oxidation \cite{Arora2021}, the search and design of 2D semiconductors with inherent resistance to oxidation can supply suitable candidate materials for practical electronic devices. In 2016, Gioele et al. observed that mechanically exfoliated 2D MoS$_2$ and MoSe$_2$ exhibit remarkable air stability, with no degradation observed after 27 and 9 days of ambient exposure, respectively \cite{Mirabelli2016}. Santosh et al. explained that the dissociative absorption of O$_2$ molecules on the MoS$_2$ surface is kinetically inhibited due to a high-energy barrier of 1.59 eV \cite{Santosh2015}. In 2017, Guo et al. \cite{Guo2017} predicted that 2D group-IV monochalcogenides (GeS, GeSe, SnS, and SnSe) as oxidation-resistant candidates with high activation energies (1.26-1.60 eV) for the O$_2$ chemisorption and robust electronic structures under the chemisorption of a moderate amount of oxygen atoms. In 2023, Daria et al. reported that exfoliated 2D GaS maintain structural integrity for more than three weeks under ambient conditions \cite{Hlushchenko2023}, which is desirable for fabricating stable devices with it.

The growth of 2D materials typically introduces various defects, including vacancies, interstitial atoms, antisites, and dislocations \cite {Komsa2012,Zhou2013,Hong2015}. These defects generate dangling bonds, introduce local defect states within the band structure, and increase the chemical reactivity of 2D materials \cite {F.Banhart2011,A.Hashimoto2004,Oezcelik2013}. For example, surface chalcogen vacancies in transition metal dichalcogenides (TMDs) and group-IV monochalcogenides substantially reduce the activation energy for oxygen dissociative adsorption \cite{Guo2017a,Gomes2016,Santosh2015}, thus accelerating their oxidation rates under ambient conditions \cite{Lee2019,Sutter2019,Higashitarumizu2018,Li2016,Groenborg2019}. Interestingly, Guo et al. reported that oxygen passivation of chalcogen vacancies in group-III monochalcogenides (MX, M = Ga, In; X = S, Se) only moderately altered their band gap and electron effective masses \cite{Guo2017a}. This finding suggests that achieving excellent oxidation resistance in 2D semiconductors requires both structural integrity and preservation of electronic properties upon oxygen interaction.

The oxidation of 2D semiconductors will generate various oxides. 2D oxide semiconductors containing the O element tend to be air-stable \cite{xie2022recent}. Among these, germanium dioxide (GeO$_2$) is an important wide-bandgap semiconductor \cite{Labed2025}. Ultra-thin GeO$_2$ films have been grown through plasma oxidation \cite{Seo2017}, magnetron sputtering \cite{Wei2024}, and chemical vapor deposition \cite{Shibayama2015}. GeO$_2$ films have shown exceptional promise for low-light-loss photonic devices \cite{smiller2020low}, advanced thin-film transistors \cite{mizoguchi2022solid}, and wide-spectrum response photodetectors \cite{liu2025high}. Recent advances have extended to two-dimensional crystalline germanium oxides. In 2019, Guo et al. proposed a 1T-type hexagonal lattice as the ground-state structure for the GeO$_2$ monolayer (ML)~\cite{Guo2019}, which was subsequently experimentally realized by Zhang et al. (2021) through controlled interfacial oxidation~\cite{Zhang2021}. GeO$_2$ ML has recently emerged as a promising 2D material due to its exceptional optoelectronic and thermal properties. With a wide band gap (3.56 eV), tightly bound excitons with a binding energy of 1.55 eV~\cite{Sozen2021}, and high thermal conductivity \cite{Wan2024}, GeO ML exhibits strong potential for application in ultraviolet (UV) photodetector and photocatalytic water splitting~\cite{Riaz2023,li2023two}.While the intrinsic properties of 2D GeO${_2}$ have been extensively studied, its environmental stability remains poorly understood. The effect of temperature and pressure on the oxidation process of 2D GeO${_2}$ is not clear. Therefore, the oxidation kinetics of 2D GeO${_2}$ calls for a microscopic-level investigation to address this research gap. 

In this work, we systematically investigated the oxidation resistance of GeO${_2}$ monolayer (ML). We first proved that both O atom and O$_2$ prefer physisorption on perfect GeO${_2}$ ML, due to the high active energy barrier for the dissociative oxidation. Perfect GeO${_2}$ ML's crystal lattice and electrical structures alter minimally before and after the absorption of oxygen. Then we predicted that the primary defect in GeO${_2}$ ML, surface O vacancy, promotes the chemisorption of O$_2$. 
The absorbed O$_2$ mainly affects the valence bands rather than the conduction bands. Interestingly, the electron effective mass of defective GeO${_2}$ can be restored to the intrinsic one by oxidation. At last, we analyzed the temperature and entropy effects on the defect formation and oxidation of GeO${_2}$ ML. Our results provides a theoretical basis for the anti-oxidation application of 2D GeO${_2}$ in a gas environment.

\section{Computational method}
The Vienna ab initio Simulation Package (VASP) \cite {b2} with the projector augmented wave (PAW) \cite {Kresse1999} pseudopotentials was utilized to perform spin-polarized first-principles calculations. We employed the Perdew-Burke-Ernzerhof (PBE) \cite {b4}  exchange-correlation functional. A vacuum layer of 18 $\mathring{\rm A}$ was introduced to eliminate spurious interactions between adjacent layers. A kinetic energy cutoff of 550 eV was applied. The convergence criteria for total energy and force were set at $10^{-5}$ eV and 0.01 eV/$\mathring{\rm A}$, respectively. The integrations of the Brillouin zone (BZ) were performed with $10 \times 10 \times 1$ Gamma-centered $\bf{k}$-mesh \cite{b5} for the primitive cell of GeO$_2$ ML.
We constructed a $5 \times 5 \times 1$ supercell to investigate the intrinsic defects and oxygen adsorption.
We considered Van der Waals (vdW) corrections using Grimme-D2 methods \cite {Chem.2006}. The reaction process was studied using the Nudged Elastic Band (NEB) method \cite {Henkelman2000}.

The formation energy $E_{f}$ of defects is defined as
\begin{equation}
	E_{f}=E_{\rm GeO_{2}+defect} - E_{\rm GeO_{2}} -n_{\rm Ge}E_{\rm Ge}  -n_{\rm O} \frac{1}{2} E_{\rm O_2}
\end{equation}
$E_{\rm GeO_{2}+defect}$, $E_{\rm GeO_{2}}$, $E_{\rm Ge}$, and $E_{\rm O_2}$ are the energies of GeO$_{2}$ ML with and without defects, the energy of a Ge atom (-4.747 eV) in bulk Ge, and the energy (-9.862 eV) of the O$_2$ molecule, respectively.

The interaction between the O$_2$ molecule and GeO${_2}$ ML is described by the binding energy ($E_{\rm bind}$) which is defined as:
\begin{equation}
	E_{\rm bind} = E_{\rm GeO{_2}+O_2} - E_{\rm GeO{_2}} - E_{\rm O_2}
\end{equation}
where $E_{\rm GeO{_2}+O_2}$, $E_{\rm GeO{_2}}$, $E_{\rm O_2}$ are the energies of GeO${_2}$ ML with an adsorbed O$_2$ molecule, the GeO${_2}$ ML, and isolated O$_2$ molecule, respectively. By definition, negative and positive $E_{\rm bind}$ means exothermic and endothermic reactions, respectively.

\begin{figure}
	\centering
	\includegraphics[width=0.45\textwidth]{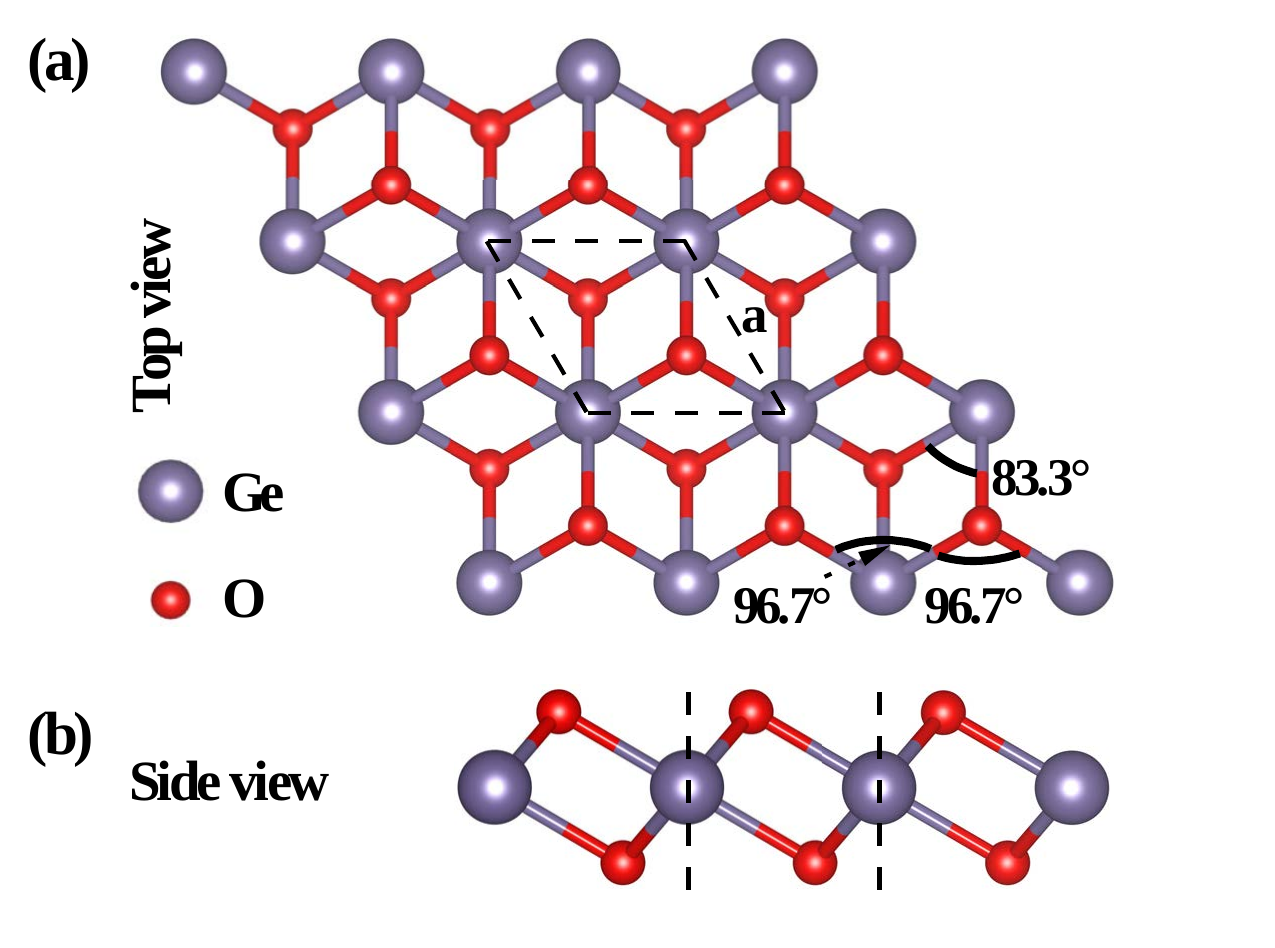}
	\caption{(a) The top and (b) side view of the GeO${_2}$ ML. The purple and red balls represent the Ge and O atoms, respectively. The primitive cell is labeled by bash lines.}
	\label{fg1}
\end{figure}

\section{Results and discussion}
\subsection{Intrinsic GeO${_2}$ monolayer}

Fig.~\ref{fg1} shows the 1T-type hexagonal crystal lattice of GeO${_2}$ ML.
The Ge layer is sandwiched between the upper and lower O layers.
Six O atoms form an octahedral configuration around the core Ge atom.
Each O atom coordinates with three neighboring Ge atoms.
GeO${_2}$ ML has the P$\overline{3}$m2 space group and the $D_{\rm3d}$ point group.
The optimized lattice constant $a$ is 2.911 \AA\ and the Ge-O bond length is 1.948 \AA.
The angle of the Ge-O-Ge bond $\angle_{\rm Ge-O-Ge}$ is 96.7$^{\circ}$.
Two types of O-Ge-O bond angle $\angle_{\rm O-Ge-O}$ are 96.7$^{\circ}$ and 83.3$^{\circ}$, respectively.
Spin-polarization calculations show that GeO${_2}$ ML is not magnetic.
GeO${_2}$ ML has an indirect band gap of 3.536 eV [see Fig. S1(a)].
The valence band maximum (VBM) is mainly contributed by the O-$p$ orbitals, while the conduction band minimum (CBM) is mainly contributed by the Ge-$s$ orbitals [see Fig. S1(b)].
These results agree with previous studies \cite{Sozen2021}, indicating the reliability of our calculations.

\begin{figure*}
	\centering
	\includegraphics[width=0.7\textwidth]{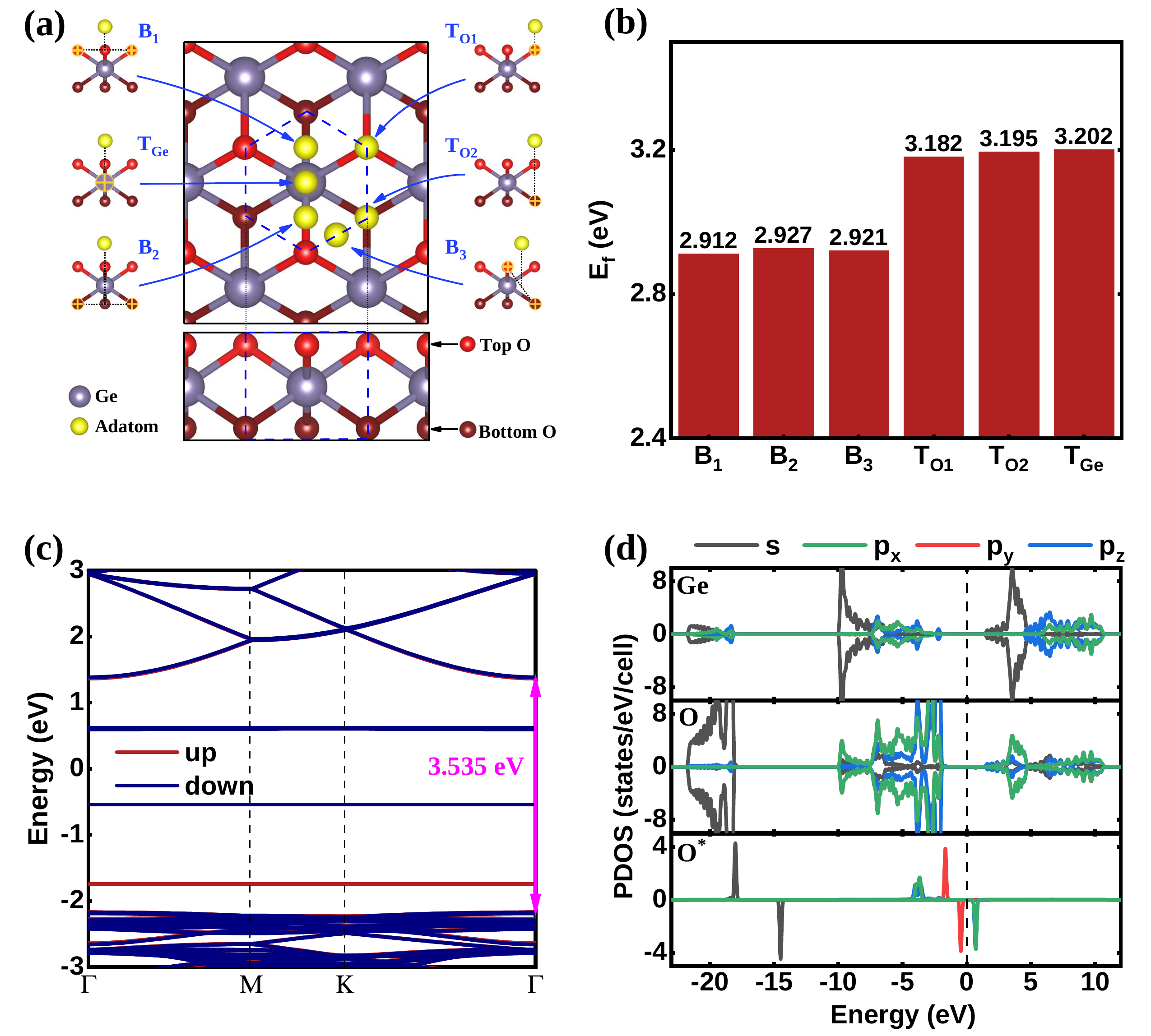}
	\caption{(a) The top and side view of six possible physisorption sites of single O atom on the GeO$_2$ ML. The O atoms in top layer and bottom layer are presented with red and dark red, respectively. The absorbed O atom is labeled by yellow.
		(b) The formation energy $E{_f}$ of different physisorption sites in O-rich conditions.
		(c) The band structure and (d) projected density of states (PDOS) of the ${\rm B}_{\rm 1}$ site.
		O and O$^{\star}$ represent the contribution of PDOS from the O atoms in GeO${_2}$ and the absorbed O atom, respectively.}
	\label{fg2}
\end{figure*}

\subsection{Oxidation of intrinsic GeO${_2}$ monolayer}
\subsubsection{Resistance to oxygen atom}

When oxygen molecules are adsorbed on the surface of 2D material, oxygen atoms can dissociate under an electron beam \cite {Naclerio2020}.
We first consider the adsorption of oxygen atoms on the surface of GeO$_2$ ML.
To simulate the physisorption of O atom, we placed an O atom on the surface of a 5$\times$5$\times$1 GeO${_2}$ supercell.
Fig.~\ref{fg2}(a) depicts possible adsorption sites on GeO${_2}$ ML, including bridged sites (B$_1$, B$_2$, B$_3$) and top sites (T$_{\rm Ge}$, T$_{\rm O1}$, T$_{\rm O2}$).
The bridged sites are on the top of the middle position between two O atoms in the top layer (labeled top-O), two O atoms in the bottom layer (labeled bottom-O), or a Ge-O bond.
The top sites represent the positions on the tops of Ge, top-O or bottom-O atoms.
The O atom in T$_{\rm Ge}$ has a larger $E_{f}$ than B$_{1}$, contrary to our first expectations.

\begin{figure*}
	\centering
	\includegraphics[width=0.7\textwidth]{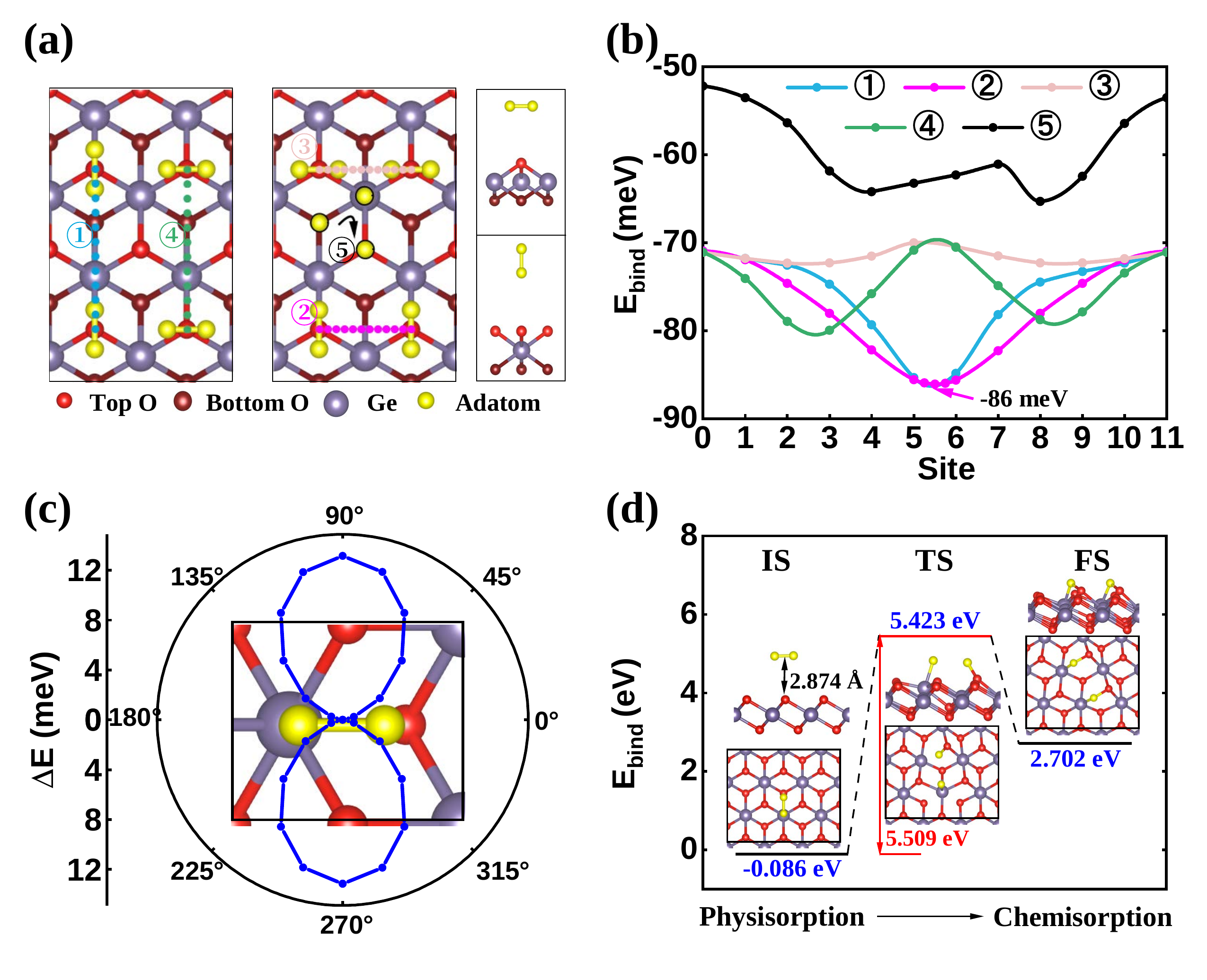}
	\caption{(a) Five diffusion paths of an O$_2$ molecule physisorbed on the GeO${_2}$ ML. The side view shows that O$_2$ molecule can be perpendicular or parallel to GeO${_2}$ ML. 
		(b) The evolution of binding energy $E_{\rm bind}$ along the diffusion paths. 
		(c) The polar coordinate diagram of the system energy as a function of the rotation angle of O$_2$ molecule in the ground-state absorbed configuration. 		
		(d) The reaction path diagram of dissociative oxidation processes on GeO${_2}$ ML. The black line segments and the red line segments respectively represent the energy levels of the physisorption and chemisorption states and the transitional state, as well as the corresponding atomic structures (top and side views). The blue numbers beside the energy levels (from left to right) respectively give the binding energies of the initial state, the transition state and the final state. The red numbers represent the activation energy (E$_a$) of the reaction. Ge atoms, substrate oxygen atoms and adsorbed oxygen atoms are respectively represented in purple, red and yellow. Some information on bond lengths and adsorption heights is shown in Fig. S2.}
	\label{fg3}
\end{figure*}

We relaxed the lattice of supercell containing defects.
Fig.~\ref{fg2}(b) shows the formation energy $E_{f}$ of single O atom in GeO${_2}$ ML.
We used various starting heights for simulating the adsorption of single O absorption to ensure the reliability of our calculations. The most stable site is the B$_{1}$ site, which is the bridge position between the Ge atom and the neighboring bottom O atom. The minor energy difference between B$_{1}$, B$_{2}$, and B$_{3}$ suggests that the O atom can diffuse on the surface of GeO$_2$ ML. The O atom in T$_{\rm Ge}$ has a higher $E_{f}$ than B$_{1}$, contrary to our initial anticipation.


We performed the Bader charge analysis and found that a tiny charge of 0.04 $e$ transfers from GeO${_2}$ ML to a physisorbed O atom. In GeO${_2}$ ML, negative top-O ions repel the physisorbed O atom, whereas positive Ge ions attract it. We adopted the point charge mode to measure the electronic potential energy: $E=k \frac{q_{1} q_{2}}{r}$. Here, $r$ is the distance between two point charges, and $k$ is the Coulomb constant. We assumed an effective charge of $2q$ and $-q$ for the Ge and O ions, respectively. The B$_{1}$ site has a vertical difference of 2.280 \AA\ from the GeO${_2}$ ML, and a distance of 2.704 \AA\ and 3.382 \AA\ from the top-O atoms and Ge atom, respectively. This O-O distance is much larger than length (1.233 Å) of typical O-O bond. The T$_{\rm Ge}$ site has a distance of 2.979 \AA\ and 3.458 \AA\ from the top-O atoms and the Ge atom, respectively. With these information, we calculated the ratio of the electronic potential energy at the $B_{1}$ and $T_{\rm Ge}$ sites to be about 0.35. The top-O atoms exert a weaker Coulomb repulsive force on the absorbed O atom at the B$_{1}$ position compared to the T$_{\rm Ge}$ site.

Fig.~\ref{fg2}(c) display the band structure of the B$_{1}$ configuration. The interaction between the absorbed O atom (labeled as O$^*$) and GeO$_2$ ML is very weak. Thus, the $p_y$ orbital in the bandgap should be treated as the atomic level of the absorbed O$^*$ atom rather than the electronic bands of the absorption system. The dispersionless bands at the energy zone of [-2 eV, 1 eV] also reflect the above discussion. We thus defined the band gap of GeO$_2$ ML as the energy difference between the valence and conduction bands whose main contribution is from GeO$_2$ ML. 
The band gap of GeO$_2$ ML of the B$_1$ configuration is 3.535 eV, nearly identical to that of intrinsic GeO$_2$ ML (3.536 eV).
The physisorbed O atom's $s$ state is deep in the valence bands, whereas the occupied $p_y$ state is within the band gap of GeO$_2$ ML [see Fig.~\ref{fg2}(d)].
The spin-up $p_x$ and $p_z$ states exhibit a broadening PDOS peak at approximately -5 eV, resulting in a local magnetic moment of 2 $\mu_B$.
The physisorbed O atom has a weak hybridization with the Ge-$p$ and O-$p$ states in GeO$_2$ ML, due to the energy overlap between the PDOS around -5 eV. 
GeO$_2$ ML's top atoms have a weak magnetic moment of 0.02 $\mu_B$ due to proximity effects.
Based on the above analysis, the physisorbed O atom weakly affects the lattice structure and electronic structure of GeO$_2$ ML. 
When the O atom approaches GeO$_2$ ML, the system's energy increases significantly, indicating a larger energy barrier for the chemisorption of the O atom in GeO${_2}$ ML, which we will discuss later.

\subsubsection{Resistance to oxygen molecule}
We explored the physisorption of the O$_2$ molecule in intrinsic GeO${_2}$ ML.
To begin, we designed diffusion pathways connecting the possible absorbed sites [see Fig.~\ref{fg3}(a)].
During the diffusion process, we maintained the orientation of the O$_2$ molecule while relaxing the vertical distance.
O$_2$ molecule is perpendicular to the GeO${_2}$ ML in path 5, while it is parallel to the GeO${_2}$ ML on other paths.

Fig.~\ref{fg3}(b) shows the evolution of the binding energy $E_{\rm bind}$ throughout several diffusion pathways.
Path 5 has a larger $E_{\rm bind}$ than other paths, showing that the O$_2$ molecule tends to align parallel rather than perpendicular to GeO${_2}$ ML.
We selected the absorption sites with the lowest $E_{\rm bind}$ for each diffusion path.
Paths 1 and 2 predicted the same stable physisorption configuration for the O$_2$ molecule.
Next, we rotated the O$_2$ molecule at the chosen absorption sites to determine the optimal orientation.
The system energy is the lowest when the O$_2$ molecule is parallel to the Ge-O bond of GeO${_2}$ ML [see Fig.~\ref{fg3}(c)].
We also rotated the O$_2$ molecule in low-energy absorbed configurations of paths 3 and 4.
We verified that these are metastable physisorbed configurations.
In the ground-state physisorbed configuration, the center of the O$_2$ molecule has a height of 2.874 \AA\ from the top layer of O [see the IS state in Fig.~\ref{fg3}(d)], and the $E_{\rm bind}$ is only 86.0 meV.

GeO$_2$ ML has a band gap of 3.538 eV after O$_2$ physisorption, which is nearly identical to its intrinsic band gap [see Fig. S3(a)]. 
The conduction bands remain highly dispersed. 
The O$_2$ molecule has a magnetic moment of 2 $\mu_B$ in the triplet state.
Peaks in PDOS are caused by the non-bonding states $s$ and $p$ of oxygen molecules [see Fig. S3(b)]. 
The presence of physisorbed O$_2$ molecules has a minimal impact on the lattice and electronic structure of GeO${_2}$ ML, according to the findings.

Next, we investigate the chemisorption of O$_2$ molecules in GeO$_2$ ML. 
O$_2$ molecule dissociates into two oxygen atoms to chemisorb on the surface of GeO$_2$ ML by overcoming a certain activation energy ($E_a$). 
As the O$_2$ molecule approaches the GeO${_2}$ ML, the O-O bond breaks [see TS in Fig.~\ref{fg3}(d)], resulting in two separate O atoms binding to monolayer. The $p$ states of the chemisorbed O atom hybridize with Ge and O atoms in an energy range of -10 to 0 eV, as shown in Fig. S3(c, d). 
The transition from physisorption to chemisorption is marked by a reaction heat of 2.788 eV. 
The transition requires an activation energy ($E_a$) of 5.509 eV, significantly higher than or comparable to the monolayers of other 2D chalcogenides, such as Ga${_2}$O${_3}$ (4.780 eV) \cite{dong2021investigations}, group-IV monochalcogenides (1.26$\sim$1.60 eV) \cite{Guo2017}, Cr${_2}$Ge${_2}$Te${_6}$ (0.48 eV) \cite{guo2022oxidation}, WY${_2}$ (T = S, Se, Te) (5.05$\sim$5.47 eV) \cite{rawat2024first}, and $\alpha$-Au${_2}$S (2.89 eV) \cite{wu2019two}.

PBE functional may fail to accurately predict the binding energy of molecular O$_2$ on 2D materials. According to Prof. Guo’s paper \cite{Guo2019a}, we adopt the experimental value of O$_2$ bonding energy ($E_{\rm bond} = -5.16$ eV~\cite{Speight2016}) instead of the DFT value to correct for the overestimated bonding energy of molecular O$_2$. The modified binding energy ($E^*_{\rm bind}$) could be redefined as:
\begin{equation}
		E^*_{\rm bind} = E_{\rm GeO{_2}+O_2} - E_{\rm GeO{_2}} - \frac{1}{2} E_{\rm O} -E_{\rm bond}
\end{equation}
where $E_{\rm GeO{_2}+O_2}$, $E_{\rm GeO{_2}}$, and $E_{\rm O}$ are the energies of the GeO$_2$ ML with an adsorbed O$_2$ molecular, the individual GeO$_2$ ML, and an isolated O atom. The modified binding energies for the initial, transition and final states are -1.491, 4.018, and 1.297 eV, respectively, which are lower than those based on using calculated energy of an O$_2$ molecular. This indicates that O$_2$ could be strongly absorbed in the GeO$_2$ ML. However, the activation energy (5.509 eV) of the oxidation process remains unchanged. Thus, the chemisorption of the O$_2$ molecule in intrinsic GeO$_2$ ML is difficult to achieve under normal conditions. Intrinsic GeO${_2}$ ML remains inert in the oxygen environment.

\subsection{Intrinsic defects in GeO${_2}$ monolayer}
In the process of growing 2D materials, defects and impurities are inevitable \cite {Komsa2012,Zhou2013,Hong2015}.
Defects can alter the oxidation process of 2D materials \cite{Guo2017a,Gomes2016,Santosh2015}.
We analyze intrinsic defects in GeO$_2$ ML, including a single vacancy of Ge (SV$_{\rm O}$) or O atom (SV$_{\rm Ge}$);
double vacancy of two Ge atoms (DV$_{\rm Ge{_2}}$), two O atoms (DV$_{\rm O{_2}}$), or a Ge-O pair (DV$_{\rm GeO}$);
anti-site defects such as replacement of a Ge atom with an O atom (Ge$_{\rm O}$) or vice versa (O$_{\rm Ge}$).
For double vacancies, we considered different distances between the vacancies, such as the nearest-neighboring (nn) and the next-nearest-neighbor (nnn) distance.

We relaxed the lattice of defective GeO$_2$ ML and showed stable defects in Fig. S4.
The O$_{\rm Ge}$ defect is not stable because the substituted Ge atom will leave GeO$_2$ ML during structural relaxation.
In the SV$_{\rm Ge}$ defect [see Fig. S3(a)], the neighboring O atoms move away from the Ge vacancy, while the neighboring Ge remains.
The length of the Ge-O bond $d\rm{_{Ge-O}}$ around the Ge vacancy becomes less than $d\rm{_{Ge-O}}$ of 1.948 \AA\ of intrinsic GeO${_2}$ ML.
The bond angle $\angle_{\rm Ge-O-Ge}$ and $\angle_{\rm O-Ge-O}$ around the Ge vacancy are larger than that of intrinsic GeO${_2}$ ML.
DV$_{\rm Ge{_2}}$ [see Fig. S4(b)] exhibit a similar lattice distortion around the Ge vacancy.
On the other side, in the SV$_{\rm O}$ defect [see Fig. S4(c)], the neighboring $d\rm{_{Ge-O}}$ is larger than that of intrinsic GeO${_2}$ ML.
The neighboring Ge atoms tend to be close to the O vacancy, while the O atoms move away from it. 
$\angle_{\rm Ge-O-Ge}$ and $\angle_{\rm O-Ge-O}$ around the SV$_{\rm O}$ defects are smaller and larger than that of intrinsic GeO${_2}$ ML, respectively.
DV$_{\rm O{_2}}$ [see Fig. S4(d)] shows larger $d\rm{_{Ge-O}}$ but smaller bond angles $\angle_{\rm Ge-O-Ge}$ or $\angle_{\rm O-Ge-O}$ around the O vacancy due to the large movement of the neighboring Ge atom.
The anti-site Ge$_{\rm O}$ show a small change in the bond length and bond angle [see Fig. S4(f)].
DV$_{\rm GeO}$, however, display a large lattice distortion around the defect [see Fig. S4(e)]. 
Both Ge and O atoms around the defect were pulled back because of a non-equivalent force. 
$d\rm{_{Ge-O}}$ around the defects becomes clear smaller than that of intrinsic GeO${_2}$ ML.

Table~\ref{table1} shows the formation energy $E_f$ and magnetic moment $M$ of various defects.
The SV$_{\rm O}$ defect exhibits the lowest $E_f$ of 4.454 eV and zero magnetic moment, followed by non-magnetic DV$_{\rm O{_2}}$ defect which has a $E_f = 8.440$ eV. Moreover, DV$_{\rm O{_2}}(nn)$ defect has a smaller $E_f $ than DV$_{\rm O{_2}}(nnn)$ under the Ge-rich condition. That means that O vacancies tend to gather together. Defects with Ge vacancies, such as SV$_{\rm Ge}$, DV$_{\rm Ge_{2(nn)}}$, and DV$_{\rm Ge_{2(nnn)}}$, have a non-zero magnetic moment and a significantly greater $E_f$ than that of O vacancies. Therefore, surface O vacancies are the main defect in GeO$_2$ ML.
In comparison, 2D SnO has a higher formation energy of 3.96 eV and 0.52 eV for the O and Sn vacancies, respectively \cite{shukla2020dft}, indicateing that intrinsic GeO${_2}$ ML can better maintain the surface structure than SnO ML.

\begin{table}[bh!] 
	\caption{The formation energy $E_f$ and magnetic moment $M$ of different defects in GeO$_2$ ML.}
	\label{table1} 
	\centering 
	\begin{tabular*}{0.45\textwidth}{@{\extracolsep{\fill}}ccc} 
		\toprule 
		{Defects} & {$E_{f}(eV)$}  & $M$($\mu_B$) \\
		\midrule 
		{${\rm SV_{Ge}}$} & 10.148  & 4 \\
		{${\rm SV_{O}}$}  & 4.454 & 0 \\
		{${\rm DV_{GeO(nn)}}$} & 9.136 & 2 \\
		{${\rm DV_{GeO(nnn)}}$} & 10.231  & 2 \\
		{${\rm DV_{O_{2}(nn)}}$}  & 8.440 & 0 \\
		{${\rm DV_{O_{2}(nnn)}}$} & 8.691 & 0 \\
		{${\rm DV_{Ge_{2}(nn)}}$} & 19.793  & 8 \\
		{${\rm DV_{Ge_{2}(nnn)}}$} & 20.159  & 8 \\
		{${\rm Ge_{O}}$} & 10.296  & 0 \\
		\bottomrule 
	\end{tabular*}
\end{table}

\begin{figure}
	\centering
	\includegraphics[width=0.5\textwidth]{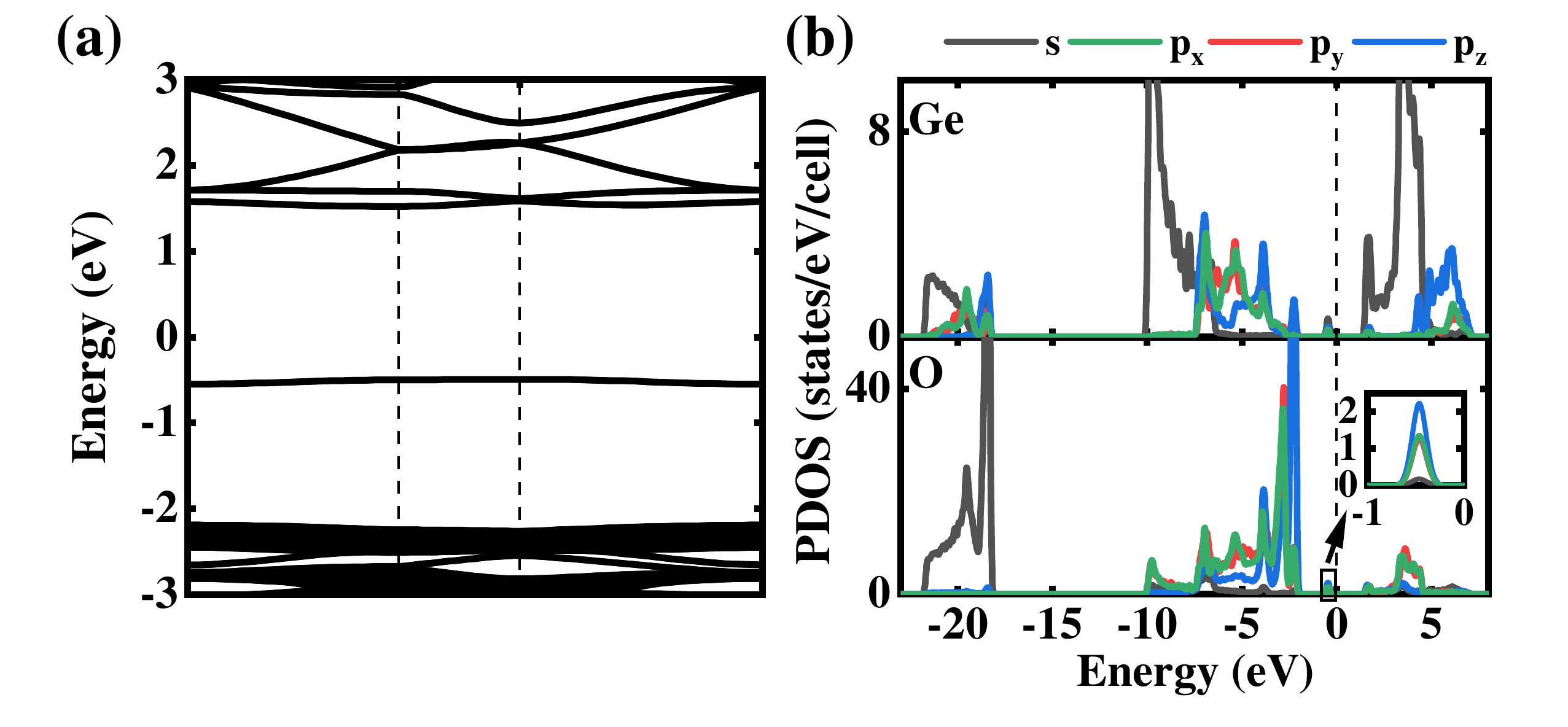}
	\caption{(a) The band structure and (b) projected density of states (PDOS) of the GeO$_2$ ML with single O vacancy (SV$_{\rm O}$ defect).}
	\label{fg4}
\end{figure}

Fig.~\ref{fg4}(a) shows the band structure of GeO$_2$ ML with the SV$_{\rm O}$ defect.
Compared to Fig. S1(a), the O vacancy introduces a defect band within the band gap of GeO$_2$ ML, reduces the band gap to 2.015 eV. 
These defect states are localized in nature and act as traps for charge carriers and recombination centers, limiting the performance of the devices \cite{Levine2019}. 
The conduction bands are more flat than that of intrinsic GeO$_2$ ML. 
Fig.~\ref{fg4}(b) shows the PDOS of the SV$_{\rm O}$ defect.
Compared to the intrinsic GeO$_2$ ML, the flattening of the conduction bands causes a peak of PDOS at 2.5 eV. 
The O-$p$ state dominates the defect state just below the Fermi level.
The dangling bond around the O vacancy will increase the chemical activity of GeO$_2$ ML. 

\subsection{Oxidation resistance of defective GeO${_2}$ monolayer}
\begin{figure}
	\centering
	\includegraphics[width=0.5\textwidth]{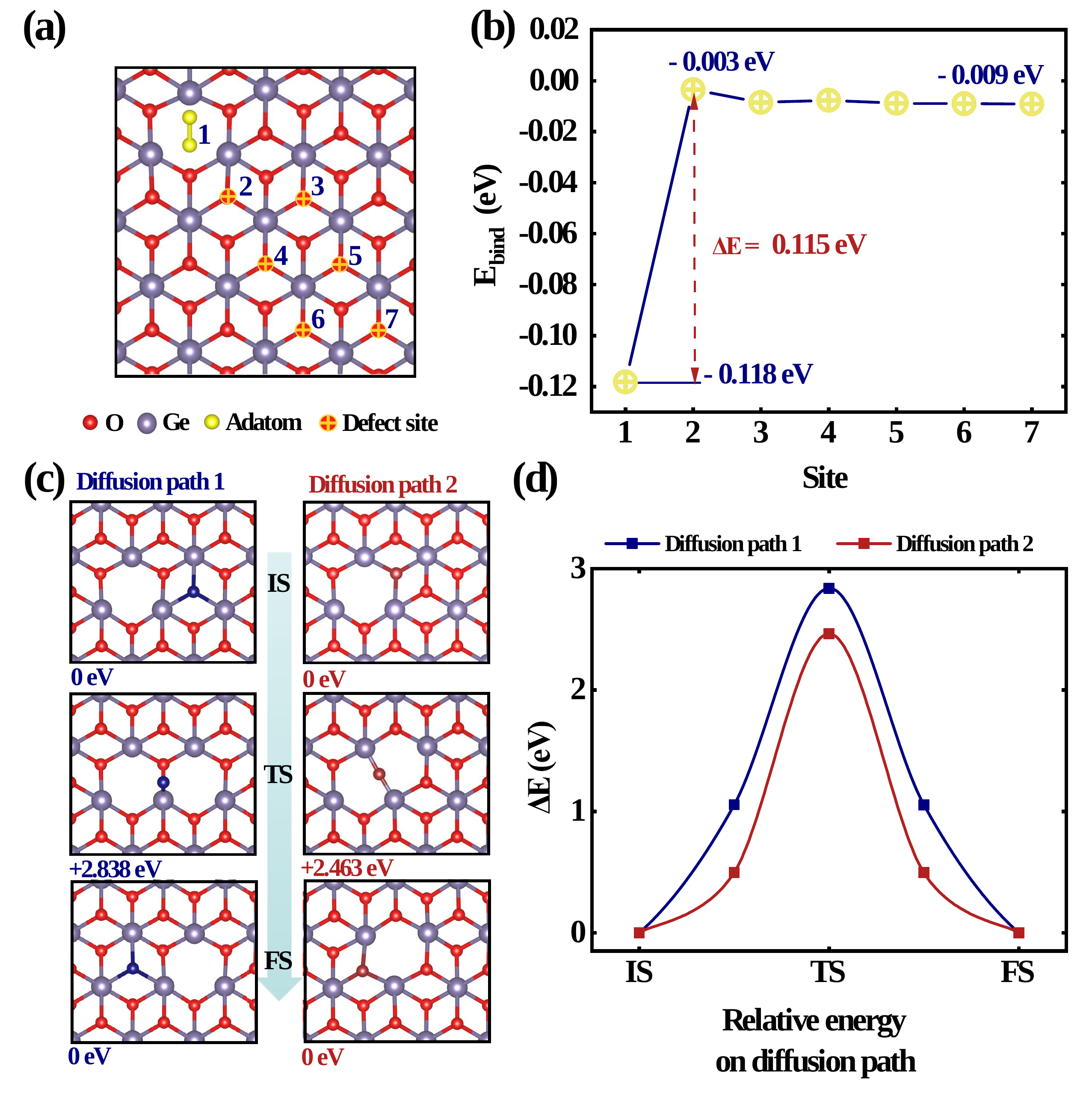}
	\caption{(a) The physisorbed O$_2$ oxygen on GeO$_2$ monolayer with O vacancy. Site 1 represent the O vacancy locates below the O$_2$ molecule. Sites 2-7 represent O vacancy with different in-plane distance from the O$_2$ molecule. (b) The system's $E_{\rm bind}$ of different sites of O vacancy. (c) The cross-layer transfer and unilateral transfer of the oxygen vacancy in the GeO${_2}$ ML. (d) Relative energy diagram along the two kinds of diffusion paths.}
	\label{fg5}
\end{figure}

We studied how the O vacancy affects the physisorption of O$_2$ molecules on GeO${_2}$ ML. We added an O$_2$ molecule to a 5$\times$5$\times$1 supercell with an O vacancy in the top-O layer [see Fig.~\ref{fg5}(a)]. 
Site 1 indicates that the O vacancy is just blowing out the physisorbed O$_2$. 
The in-plane distance between the O vacancy and absorbed O$_2$ increases from site 2 to 7. 
We allow the O$_2$ molecule to adjust itself and find a stable configuration. 
Fig.~\ref{fg5}(b) shows how the energy of the system varies based on the distance between the vacancy and O$_2$ molecules. Site 1 has the lowest energy, indicating that O$_2$ molecules choose to approach the O vacancy rather than move away from it. 

\begin{figure*}
	\centering
	\includegraphics[width=0.95\textwidth]{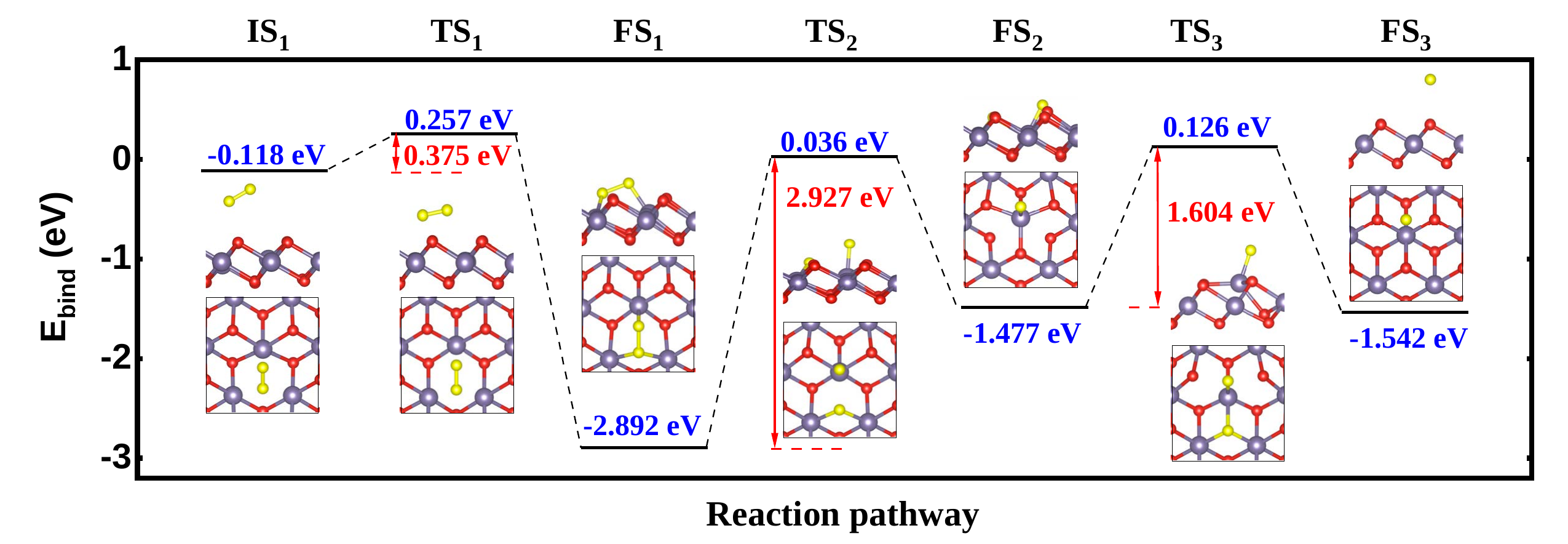}
	\caption{The transition from physisorption to chemisorption, then to dissociative oxidation for an O$_2$ molecule on the defective GeO${_2}$ ML. The black and red line segments represent the energy levels of the initial/final states and the transitional state, respectively.  The corresponding atomic structures (top and side views) are displayed. The blue numbers indicate the binding energies of the different states. The red numbers represent the activation energy (E$_a$) of the reaction. Ge and O atoms in GeO$_2$ ML are represented in purple and red, respectively. The adsorbed oxygen atoms are displayed by yellow balls. The information on bond lengths and adsorption heights is shown in Fig S6.
       }
	\label{fg6}
\end{figure*}

Compared to physisorption on intrinsic GeO$_2$ ML [see Fig.~\ref{fg3}(c)], O$_2$ molecule goes to the center of the hexagonal ring and locates at the top of the O vacancy.  
The IS$_1$ state in Fig.~\ref{fg6} shows the side view of a stable physisorption configuration.   
The $E_{\rm bind}$ of an O$_2$ molecule physisorbed in defective GeO$_2$ ML is -0.118 eV, which is higher than that of intrinsic GeO$_2$ ML (-0.086 eV).
O vacancy promotes the physisorption of the O$_2$ molecule on GeO$_2$ ML.  
O$_2$ must overcome a small energy barrier of 0.006 eV as it migrates to the near-O vacancy along the sheet [see Fig.~\ref{fg5}(b)].
On the other hand, the O$_2$ molecule crosses a large energy barrier of 0.115 eV to leave the O vacancy.

Compared with the fast migration of the O$_2$ molecule in GeO$_2$ ML, the diffusion of the O vacancy is slower. 
We designed two paths for movement of the O vacancy, namely unilateral transfer and cross-layer transfer [see Fig.~\ref{fg5}(c)].
In the first path, the next-neighboring O atom crosses the Ge-O bond and fills the O vacancy, leaving another O vacancy in its original position. 
The second path involves the unilateral movement of the nearest-neighboring O atom, resulting in the diffusion of the O vacancy. 
The second path has an energy barrier of 2.463 eV, which is lower than that of the first path's barrier of 2.838 ev [see Fig.~\ref{fg5}(d)], but significantly larger than that of migration of O$_2$ molecule (0.006 eV).  

We simulate the transition from physisorption to chemisorption of an O$_2$ molecule on the defective GeO${_2}$ ML.
Initially, the physisorbed O$_2$ molecule stays on top of the O vacancy with a height of about 2.094 \AA\ [see initial state (IS${_1}$) in Fig.~\ref{fg6}].
As the O$_2$ molecule approaches GeO${_2}$ ML, the length of the O-O bond increases from 1.234 \AA\ to 1.436 \AA.
The transition state (TS${_1}$) has a $E_{\rm bind}$ of 0.257 eV.
In the final state (FS$_1$), O$_2$ molecule occupies the O vacancy and forms chemical bonds with three neighboring Ge atoms.
FS$_1$ state has no magnetic moment.
The $E_{\rm bind}$ of the FS$_1$ state is -2.892 eV.
The reaction heat of the state IS$_1$$\to$$FS_1$ was -2.774 eV, indicating an exothermic reaction.

In this process, O$_2$ molecule overcomes an activation energy ($E_a$) of 0.375 eV, which is lower than that of phosphorene (0.54 eV) \cite {Ziletti2015} and defective MoS$_2$ ML (0.8 eV) \cite {KC2015}, but larger than defective group-III monochalcogenide (0.26-0.36 eV) \cite{Guo2017a}. Moreover, the $E_a$ of defective GeO$_2$ ML is significantly lower than that of the intrinsic layer (5.509 eV). The defect-induced decrease in the $E_a$ has been observed in MoS$_2$ (1.59$\to$0.80 eV) \cite {KC2015} and GaS (3.11$\to$0.36 eV) \cite{Guo2017a} ML. Furthermore, the $E_a$ value is less than the 0.910 eV critical energy barrier for reaction at room temperature \cite {Jiang2018}. Therefore, the O$_2$ molecule can chemisorb on the defective GeO${_2}$ ML at ambient temperature. 

Moreover, we have checked that the other defects have not only a higher formation energy but also a higher activation energy ($E_a$) of O$_2$ chemisorption than the O vacancy. Taking the Ge-O pair defect as an example,  Fig. S5 shows that its $E_a$ for the O$_2$ chemisorption is 1.626 eV, much higher than that of the O defect (0.375 eV). Thus, the creation and oxidation of Ge-O pair defect is difficult to occur.

In addition, we investigated the dissociative oxidation process, in which the O-O bond of the O$_2$ molecule breaks. 
An oxygen atom occupies the O vacancy, while the other O atom bonds with a Ge atom nearby [see finial state 2 (FS${_2}$) in Fig.~\ref{fg6}]. 
FS$_2$ state is not magnetic.
This process goes through the transition state TS${_2}$, yielding a high activation energy $E_a$ of 2.927 eV.
Moreover, the binding energy $E_{\rm bind}$ of FS${_2}$ state is 1.415 eV higher than that of FS${_1}$ state.
Therefore, it is difficult for the dissociative oxidation process to occur at room temperature.
The FS$_1$ state is the preferred chemisorption configuration of an O$_2$ molecule on GeO${_2}$ ML with the O vacancies.

\begin{figure}
	\centering
	\includegraphics[width=0.5\textwidth]{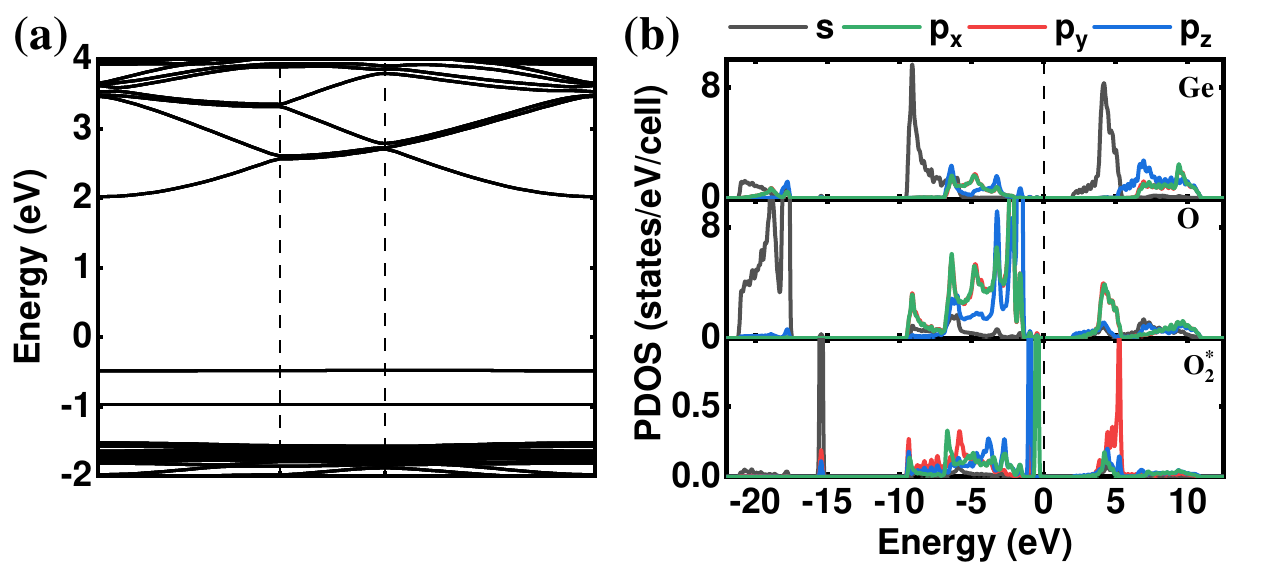}
	\caption{(a) The electronic bands and (b) projected density of states (PDOS) of FS${\rm _{1}}$ state. The "Ge" and "O" present the PDOS from the Ge and O atoms on GeO${_2}$ ML, respectively. "$\rm O{_2^\star}$" stands for PDOS of absorbed O$_2$.}
	\label{fg7}
\end{figure}

To show how chemisorbed O$_2$ affects the electrical structure of defective GeO$_2$ ML, Fig.~\ref{fg7}(a) and (b) displays the band structure and PDOS of the FS${\rm _{1}}$ state. 
Stronger hybridization between O-dimer and Ge atoms leads to a lower band gap (2.498 eV) in the FS${\rm _{1}}$ state compared to the intrinsic GeO$_2$ ML. The oxygen molecule's $p$ orbitals dominate electron states near the VBM. The FS${\rm _{1}}$ state's conduction band exhibits greater diversity compared to the SV$_{\rm O}$ structure. Table~\ref{t4} displays the carrier effective masses for both the ideal GeO${_2}$ ML and the FS${\rm _{1}}$ state. 
The intrinsic GeO${_2}$ ML has a small electron effective mass ($m_e$) of 0.353 $m{_0}$ and a large hole effective mass ($m_h$) of 3.384 $m{_0}$. 
The FS${\rm _{1}}$ structure has a $m_e$ of 0.364 $m{_0}$, which is almost that of intrinsic GeO${_2}$ ML.
Exposing defective GeO${_2}$ ML to dry oxygen restores its electron transport characteristics. 

The similar phenomenon has reported in $\beta$-In$_2$S$_3$. The exposure to the environment leads to a large number of S vacancies in $\beta$-In$_2$S$_3$, affecting the charge transport. After continuous exposure (O atoms occupying S vacancies), the carrier mobility increased nearly three times compared to the defect layer (0.023$\to$0.072 cm$^2$V$^{-1}$S$^{-1}$). Therefore, oxidation serves as a tool to repair the electron transport performance of chalcogenides \cite {huang2021defects}. 

On the other band, the $m_h$ of the FS${\rm _{1}}$ structure rises significantly to 11.809 $m{_0}$, highlighting the profound impact of defects and oxygen exposure on hole mobility. 
Therefore, the introduction of O vacancy and chemisorbed O${_2}$ act as traps or scattering centers for holes rather than electrons.  

Based on the above analysis, we predict that n-type GeO${_2}$—where the Fermi level lies close to the conduction band — exhibits electron transport properties that are robust against oxidation. Furthermore, since the electron effective mass is smaller than that of holes, the n-type 2D GeO${_2}$ is expected to be more suitable for practical electronic devices compared to its p-type counterpart.

\begin{table}[bh!] 
	\small 
	\caption{The band gap $E_{g}$ (eV) and carrier effective mass (in the unit of free electron mass $m_0$) of the intrinsic GeO$_2$ monolayer and FS$_1$ state in fig.~\ref{fg6}(a).}
	\label{t4} 
	\centering 
	\begin{tabular*}{0.4\textwidth}{@{\extracolsep{\fill}}cccc} 
		\toprule 
		& $E_{g}$ (eV) & $m_{e}$ ($m_0$) & $m_{h}$ ($m_0$) \\ %
		\midrule 
		intrinsic GeO$_2$ & 3.536 & 0.353 & 3.384 \\
		FS$_1$ & 2.498 & 0.364 & 11.809 \\
		\bottomrule 
	\end{tabular*}
\end{table}

In the FS${_2}$ state, the dissociative O atom chemisorbs with GeO$_2$ ML and forms a Ge-O bond of 1.821 $\mathring{\rm A}$.  
The FS${\rm _{2}}$ state has an impurity band just below the conduction bands of GeO$_2$ ML [see Fig. S6], which have a great impact on the electron transport properties of GeO$_2$ ML. 
We tested whether the dissociative O atom can easily detach from the surface of GeO${_2}$ ML, and restore the physisorption.
Fig.~\ref{fg6} shows that the transition from chemisorption to physisorption of O atom results in a small reaction heat of -0.070 eV.
However, the whole process requires an activation energy of 1.604 eV. 
This result shows that the defective GeO${_2}$ ML can restore the intrinsic structure after oxidation, but the desorption of the chemisorbed O atom is slow at room temperature.

\subsection{Oxidation kinetics of GeO${_2}$ ML}
Among the oxidation of GeO$_2$ ML, we observed a cyclic evolution process under low-concentration oxygen environment, which consists of four endothermic reactions.
\begin{align}
\text{I:}  & \quad \ce{GeO2 -> SV_O + O} \label{1}\\
\text{II:} & \quad \ce{SV_O + O2 -> FS_1} \label{2}\\
\text{III:} & \quad \ce{FS_1 -> FS_2} \label{3}\\
\text{IV:} & \quad \ce{FS_2 -> GeO2 + O \label{4}}
\end{align}

In the step I, intrinsic GeO$_2$ ML produces an oxygen vacancy.
In step II, O$_2$ molecule occupies the surface O vacancy as an O dimer. 
The electron transport characteristic of GeO$_2$ ML was assembled to a intrinsic layer, while the hole transport properties varied. Then, O$_2$ molecule is split into two O atoms in step III: one O atom is bonded to the Ge atom alone, while the other atoms remain in the O vacancy. At last, in step IV, the chemisorbed O atom desorbs from the GeO$_2$ ML. We denote the activation energy for each step as $E_{a1}$, $E_{a2}$, $E_{a3}$, and $E_{a4}$, respectively.
 
We analyzed the influence of temperature and entropy on the reaction thermodynamics in above four reactions. At finite temperatures, the activation energy ($E_a$) is determined by the Gibbs free energy change of transition state and initial state \cite {intayot2025unveiling}: 
\begin{equation}
	\Delta G = \Delta E + \Delta (pV) + \Delta E_{\mathrm{ZPE}} - T\Delta S
\end{equation}
where $\Delta E$ is the total energy difference obtained from DFT calculations at 0 K. 
The $\Delta E$ of each step has been labeled in Fig. S7 and Fig.~\ref{fg6}. 
$p$ is the pressure, $V$ is the volume, $\Delta E_{\mathrm{ZPE}}$ is the vibration zero-point energy correction, $T$ is the temperature. $\Delta S$ represents the entropy change $\Delta$S = $\Delta$S$_{vib}$ + $\Delta$S$_{rot}$ + $\Delta$S$_{trans}$ which contains the contribution from the vibration $\Delta$S$_{vib}$, rotation $\Delta$S$_{rot}$, and $\Delta$S$_{trans}$ translation modes, respectively. Moreover, when the number of moles (labeled as n) of a gas change during the reaction, the Gibbs free energy should included $\Delta (pV) = \Delta nRT$, where $R$ is the molar gas constant. The detailed information can be seen the part 8 of supplementary materials. 

To display the effect of temperature on the oxidation kinetics of O$_2$ chemisorption, we estimated the approximate reaction time $\tau$ of a chemical process with the Arrhenius equation:
\begin{equation}
	\tau=\frac{1}{\nu e^{\left(\frac{-E_{a}}{k_{B} T}\right)}}
\end{equation}
Here, $E_a$ represents the activation energy of the reaction, $k_{B}$ is the Boltzmann constant, $\nu$ is the trial frequency which can be estimated as $10^{13}\ \mathrm{Hz}$ for the oxidation of 2D materials \cite {he2018interaction,poldorn2020theoretical}. 

In step I, during the creation of an O vacancy, the energy difference $\Delta E $ is 7.583 eV (see Fig. S7). We only considered the $\Delta S$ = $\Delta S_{trans}$ for a single O atom in the free energy calculations. In a ambient condition with $T  = 298.15$ K and $p = 1$ atm, the activation energy ($E_{a1}$) is 7.176 eV. 
As temperature increases, the $\Delta S_{trans}$ arises and $E_{a1}$ decreases.    
At a high energy of 1000 K, we got $E_{a1} = 5.958$ eV. The reaction time $\tau$ decreases as temperature arises (see Table.~\ref{table3}). 
So the O vacancies trend to be formed at high temperatures. 
In the experiment, At 600 $^\circ C$ and 1atm atmospheric pressure, the oxygen vacancies in the GeO$_2$ thin sheet were observed \cite{DaSilva2012}. Moreover, though 2D Ga$_2$O$_3$ has a large activation energy of 4.78 eV \cite{dong2021investigations}, the oxygen vacancies in Ga$_2$O$_3$ thin films were observed at 1000 $^\circ C$ \cite{AlGhaithi2025}.

\begin{figure}
	\centering
	\includegraphics[width=0.45\textwidth]{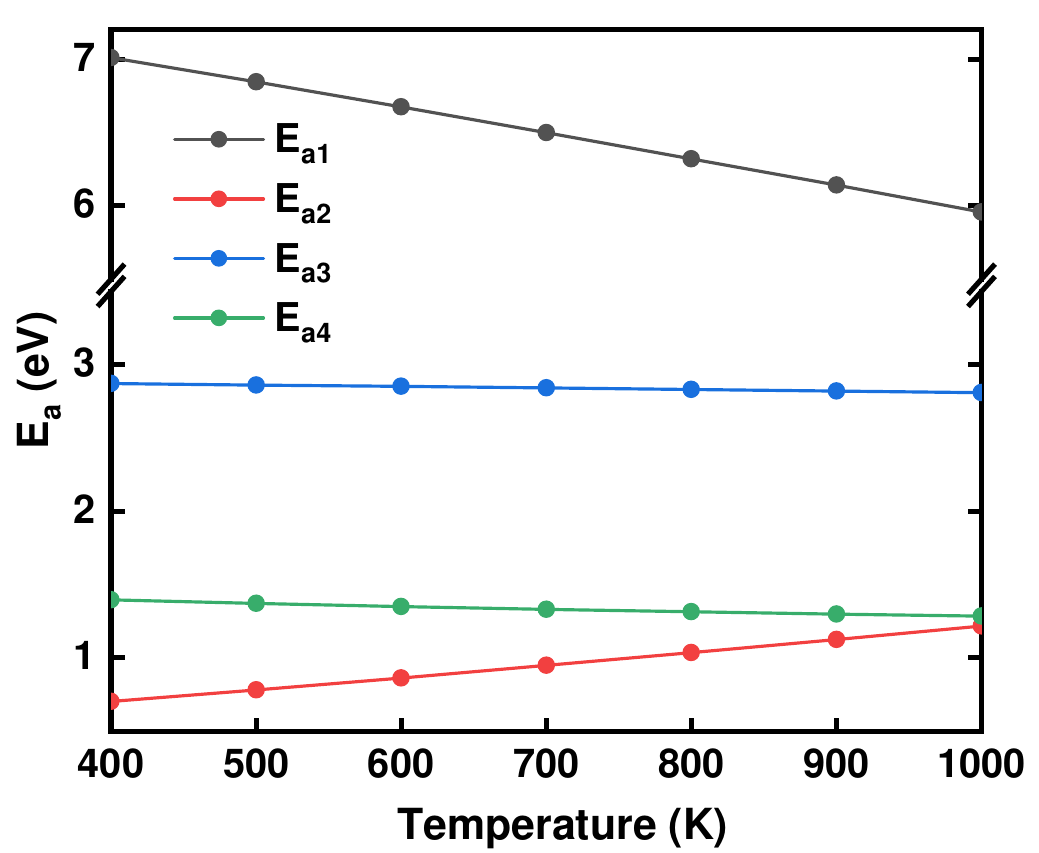}
	\caption{The evolution graph of the activation energy ($E_{a}$, $E_{a1}$, $E_{a2}$, and $E_{a3}$) of the oxidation process of defect GeO$_2$ ML under 1 atm air pressure with temperature.}
	\label{fg8}
\end{figure}

In steps II to IV under a constant pressure of $p = 1$ atm, Fig.~\ref{fg8} display the variation of activation energy ($E_{a2}$, $E_{a3}$, and $E_{a4}$) with temperature.
In step II that O$_2$ molecule chemisorbs on defective GeO$_2$ ML, our calculations show that the corresponding $\Delta S$ is negative. 
At a temperature of 400 K, the estimated reaction time $\tau$ for the chemical adsorption of O$_2$ on defect GeO$_2$ ML with O$_2$ (step~$\mathrm{II}$) was $7.05 \times 10^{-5}$ s. 
As the temperature rises, the vibrational motion of O$_2$ molecules becomes more intense. 
The corresponding increase in entropy is the primary factor driving the growth in the activation energy $E_{a2}$ for O$_2$ chemisorption at elevated temperatures.
However, the reaction time $\tau$ still decreases (see Table.~\ref{table3}). Therefore, defective GeO$_2$ ML become more vulnerable to oxidation at a higher temperature. 

In contrast, steps~\text{III} and~\text{IV}, which involve chemical bonding of oxygen with defective GeO$_2$ ML, show a very slight decrease of about 0.1 eV in their activation energy ($E_{a3}$, and $E_{a4}$) within the temperature ranging from 400 to 1000 K. These observations demonstrate that thermal effects predominantly influence the adsorption dynamics of O$_2$. 
The oxidation characteristics of structurally stabilized reactants and products in steps~\text{III} and \text{IV} remain largely invariant under thermal perturbation. Table.~\ref{table3} shows that increasing the temperature can effectively speed up the reaction rate of steps~\text{III} and~\text{IV}.

\begin{table}[ht]
	\centering
	\caption{Estimated reaction times (in s) for defect formation (step \text{I}), oxidation (step \text{II}), and reduction processes (step \text{III} and \text{IV}) in GeO$_2$ ML at selected temperatures.}
	\label{table3}
	\begin{tabular}{S[table-format=4.0] *{4}{c}}
		\toprule
		{T (\si{K})} & II & III & IV \\ 
		\midrule
		400    & $7.0\times10^{-5}$  & $1.5\times10^{23}$  & $3.9\times10^{4}$   \\
		500    & $7.5\times10^{-6}$  & $7.0\times10^{15}$  & $6.9\times10^{0}$                 \\
		600   & $1.8\times10^{-6}$  & $9.0\times10^{10}$  & $2.2\times10^{-2}$  \\
		700    & $6.8\times10^{-7}$  & $2.9\times10^{7}$   & $3.9\times10^{-4}$  \\
		800    & $3.4\times10^{-7}$  & $6.9\times10^{4}$   & $1.9\times10^{-5}$  \\
		900    & $2.0\times10^{-7}$  & $6.2\times10^{2}$   & $1.9\times10^{-6}$  \\
		1000  & $1.4\times10^{-7}$  & $1.4\times10^{1}$   & $3.0\times10^{-7}$  \\
		\bottomrule
	\end{tabular}
\end{table}

During the oxidation of defective GeO$_2$ ML (steps II-IV), step III, in which the decomposition of O$_2$ dimer into O atoms, is the rate limiting step. As the temperature increases, the reaction in step~$\mathrm{III}$ becomes possible. We noticed that the oxygen molecules on Au/Ag clusters undergo dissociation unless the temperature is greater than 700 K \cite {poldorn2020theoretical}. Meanwhile, Table.~\ref{table3} shows that the reaction rate of step II is much larger than that of step I, due to the large $E_{a1}$. According to the aforementioned analysis, the electron transport of defective GeO$_2$ ML can be restored by oxidation in the environment.     

At last, we discussed the influence of the partial pressure of oxygen on the oxidation mechanism of GeO$_2$ ML. 
The partial pressure of oxygen can adjusting the chemical potential of oxygen molecules [see Supplementary Materials, Part 8]. The increase in partial pressure of oxygen will arise the formation energy of O vacancy in step I, thereby hindering its formation. In comparison, the pressure of oxygen does not affect the activation energy in the oxidation process (step II-IV). 

\section{Conclusions}
In this study, we used first-principles calculations to investigate the oxidation resistance of GeO${_2}$ monolayer (ML). In intrinsic GeO${_2}$ ML, surface negative O ions have a natural repulsion force on the ambient oxygen molecules, as evidenced by the high activation energy of 5.509 eV for the dissociative oxidation process. Surface O vacancy is the primary defect in GeO${_2}$ ML. The in-plane diffusion of O vacancy is much slower than the migration of O$_2$ molecules on the surface of GeO$_2$ ML. Defective GeO$_2$ ML with surface oxygen vacancies becomes vulnerable to O$_2$, with a low activation energy of 0.375 eV. The O$_2$ molecule prefers to occupy the O vacancy as an O dimer rather than atomic O, due to a high activation energy of 1.604 eV for the further O$_2$ molecule splitting. Meanwhile, the conduction bands and electron effective mass nearly return to that of intrinsic GeO${_2}$, while hole transport properties alter. This means oxygen molecules in the air can restore the electron transport properties of defective GeO${_2}$. With the free energy calculations, we further proved that the high O$_2$ pressure hinders the formation of the O vacancy, while high temperature increases the oxidation rate in GeO$_2$ ML. 
These results provide theoretical insights to understand the oxidation process of 2D GeO$_2$ under realistic environmental conditions, support the practical employment of 2D GeO$_2$ in flexible electronics, UV photodetector, and catalytic systems by mitigating performance degradation caused by oxidation.
In addition, our theoretical study is limited to dry oxygen environment. Future work should explore the integral effect of O2 and water molecules on the environmental stability of 2D oxides.






\printcredits

\section*{Declaration of competing interest }
The authors declare no conflict of interest.

\section*{Data Availability}
Data will be made available on request

\section*{Acknowledgements}
This work was supported by National Natural Science Foundation of China (No. 11904313),
the Project of Hebei Educational Department, China (No. BJ2020015),
the Natural Science Foundation of Hebei Province (No. A2022203006),
the Doctor Foundation Project of Yanshan University (No. BL19008).
Innovation Capability Improvement Project of Hebei province (No. 22567605H).
The numerical calculations have been done in the High Performance Computing Center of Yanshan University.

\section*{Supplemental Material}
See Supplemental Material for further details, which includes the electronic structure of intrinsic GeO$_2$ ML, prefect GeO$_2$ ML with physisorption of O$_2$ molecule, prefect GeO$_2$ ML with the chemisorption of O$_2$ molecule. 
the structure around various defects. The electronic structure of FS$_2$ state in Fig. 6(a).
The active energy barrier of formation of surface O vacancy. 

\bibliographystyle{rsc}

\bibliography{cas-dc-template}

\providecommand*{\mcitethebibliography}{\thebibliography}
\csname @ifundefined\endcsname{endmcitethebibliography}
{\let\endmcitethebibliography\endthebibliography}{}
\begin{mcitethebibliography}{65}
\providecommand*{\natexlab}[1]{#1}
\providecommand*{\mciteSetBstSublistMode}[1]{}
\providecommand*{\mciteSetBstMaxWidthForm}[2]{}
\providecommand*{\mciteBstWouldAddEndPuncttrue}
  {\def\EndOfBibitem{\unskip.}}
\providecommand*{\mciteBstWouldAddEndPunctfalse}
  {\let\EndOfBibitem\relax}
\providecommand*{\mciteSetBstMidEndSepPunct}[3]{}
\providecommand*{\mciteSetBstSublistLabelBeginEnd}[3]{}
\providecommand*{\EndOfBibitem}{}
\mciteSetBstSublistMode{f}
\mciteSetBstMaxWidthForm{subitem}
{(\emph{\alph{mcitesubitemcount}})}
\mciteSetBstSublistLabelBeginEnd{\mcitemaxwidthsubitemform\space}
{\relax}{\relax}

\bibitem[Chhowalla \emph{et~al.}(2016)Chhowalla, Jena, and
  Zhang]{Chhowalla2016}
M.~Chhowalla, D.~Jena and H.~Zhang, \emph{Nat. Mater.}, 2016, \textbf{1},
  16052\relax
\mciteBstWouldAddEndPuncttrue
\mciteSetBstMidEndSepPunct{\mcitedefaultmidpunct}
{\mcitedefaultendpunct}{\mcitedefaultseppunct}\relax
\EndOfBibitem
\bibitem[Sheng \emph{et~al.}(2023)Sheng, Dong, Zhu, Wang, Chen, Xia, Xu, Zhou,
  Wan, and Bao]{Sheng2023}
C.~Sheng, X.~Dong, Y.~Zhu, X.~Wang, X.~Chen, Y.~Xia, Z.~Xu, P.~Zhou, J.~Wan and
  W.~Bao, \emph{Adv. Funct. Mater.}, 2023, \textbf{33}, 2304778\relax
\mciteBstWouldAddEndPuncttrue
\mciteSetBstMidEndSepPunct{\mcitedefaultmidpunct}
{\mcitedefaultendpunct}{\mcitedefaultseppunct}\relax
\EndOfBibitem
\bibitem[Kim \emph{et~al.}(2024)Kim, Kwon, Ryu, Kim, Kim, Lee, Lee, Seo, Han,
  Suh, Kim, Song, Lee, Seol, and Kim]{Kim2024}
K.~S. Kim, J.~Kwon, H.~Ryu, C.~Kim, H.~Kim, E.-K. Lee, D.~Lee, S.~Seo, N.~M.
  Han, J.~M. Suh, J.~Kim, M.-K. Song, S.~Lee, M.~Seol and J.~Kim, \emph{Nat.
  Nanotechnol.}, 2024, \textbf{19}, 895--906\relax
\mciteBstWouldAddEndPuncttrue
\mciteSetBstMidEndSepPunct{\mcitedefaultmidpunct}
{\mcitedefaultendpunct}{\mcitedefaultseppunct}\relax
\EndOfBibitem
\bibitem[Yang \emph{et~al.}(2022)Yang, Liu, Dong, Shen, Pan, Wang, Tang, Dai,
  Wu, Jin,\emph{et~al.}]{Yang2022}
J.~Yang, X.~Liu, Q.~Dong, Y.~Shen, Y.~Pan, Z.~Wang, K.~Tang, X.~Dai, R.~Wu,
  Y.~Jin \emph{et~al.}, \emph{Chin. Chem. Lett.}, 2022, \textbf{33},
  177--185\relax
\mciteBstWouldAddEndPuncttrue
\mciteSetBstMidEndSepPunct{\mcitedefaultmidpunct}
{\mcitedefaultendpunct}{\mcitedefaultseppunct}\relax
\EndOfBibitem
\bibitem[Li \emph{et~al.}(2019)Li, Zhou, Shi, Chen, and Wang]{Li2019}
Q.~Li, Q.~Zhou, L.~Shi, Q.~Chen and J.~Wang, \emph{J. Mater. Chem. A}, 2019,
  \textbf{7}, 4291--4312\relax
\mciteBstWouldAddEndPuncttrue
\mciteSetBstMidEndSepPunct{\mcitedefaultmidpunct}
{\mcitedefaultendpunct}{\mcitedefaultseppunct}\relax
\EndOfBibitem
\bibitem[Yang and Hao(2019)]{Yang2019}
Z.~Yang and J.~Hao, \emph{Adv. Mater. Technol.}, 2019, \textbf{4},
  1900108\relax
\mciteBstWouldAddEndPuncttrue
\mciteSetBstMidEndSepPunct{\mcitedefaultmidpunct}
{\mcitedefaultendpunct}{\mcitedefaultseppunct}\relax
\EndOfBibitem
\bibitem[Wu \emph{et~al.}(2021)Wu, Lyu, Zhang, Ding, Zheng, Yang, Lau, Chen,
  and Hao]{Wu2021}
Z.~Wu, Y.~Lyu, Y.~Zhang, R.~Ding, B.~Zheng, Z.~Yang, S.~P. Lau, X.~H. Chen and
  J.~Hao, \emph{Nat. Mater.}, 2021, \textbf{20}, 1203--1209\relax
\mciteBstWouldAddEndPuncttrue
\mciteSetBstMidEndSepPunct{\mcitedefaultmidpunct}
{\mcitedefaultendpunct}{\mcitedefaultseppunct}\relax
\EndOfBibitem
\bibitem[Wells \emph{et~al.}(2018)Wells, Henning, Gish, Sangwan, Lauhon, and
  Hersam]{Wells2018}
S.~Wells, A.~Henning, J.~T. Gish, V.~K. Sangwan, L.~J. Lauhon and M.~C. Hersam,
  \emph{Nano Lett.}, 2018, \textbf{18}, 7876--7882\relax
\mciteBstWouldAddEndPuncttrue
\mciteSetBstMidEndSepPunct{\mcitedefaultmidpunct}
{\mcitedefaultendpunct}{\mcitedefaultseppunct}\relax
\EndOfBibitem
\bibitem[Ahmed \emph{et~al.}(2017)Ahmed, Balendhran, Karim, Mayes, Field,
  Ramanathan, Singh, Bansal, Sriram, Bhaskaran, and Walia]{Ahmed2017}
T.~Ahmed, S.~Balendhran, M.~N. Karim, E.~L.~H. Mayes, M.~R. Field,
  R.~Ramanathan, M.~Singh, V.~Bansal, S.~Sriram, M.~Bhaskaran and S.~Walia,
  \emph{npj 2D Mater. Appl.}, 2017, \textbf{1}, 18\relax
\mciteBstWouldAddEndPuncttrue
\mciteSetBstMidEndSepPunct{\mcitedefaultmidpunct}
{\mcitedefaultendpunct}{\mcitedefaultseppunct}\relax
\EndOfBibitem
\bibitem[Arora and Erbe(2021)]{Arora2021}
H.~Arora and A.~Erbe, \emph{InfoMat}, 2021, \textbf{3}, 662--693\relax
\mciteBstWouldAddEndPuncttrue
\mciteSetBstMidEndSepPunct{\mcitedefaultmidpunct}
{\mcitedefaultendpunct}{\mcitedefaultseppunct}\relax
\EndOfBibitem
\bibitem[Mirabelli \emph{et~al.}(2016)Mirabelli, McGeough, Schmidt, McCarthy,
  Monaghan, Povey, McCarthy, Gity, Nagle, Hughes, Cafolla, Hurley, and
  Duffy]{Mirabelli2016}
G.~Mirabelli, C.~McGeough, M.~Schmidt, E.~K. McCarthy, S.~Monaghan, I.~M.
  Povey, M.~McCarthy, F.~Gity, R.~Nagle, G.~Hughes, A.~Cafolla, P.~K. Hurley
  and R.~Duffy, \emph{J. Appl. Phys.}, 2016, \textbf{120}, 125102\relax
\mciteBstWouldAddEndPuncttrue
\mciteSetBstMidEndSepPunct{\mcitedefaultmidpunct}
{\mcitedefaultendpunct}{\mcitedefaultseppunct}\relax
\EndOfBibitem
\bibitem[KC \emph{et~al.}(2015)KC, Longo, Wallace, and Cho]{Santosh2015}
S.~KC, R.~C. Longo, R.~M. Wallace and K.~Cho, \emph{J. Appl. Phys.}, 2015,
  \textbf{117}, 135301\relax
\mciteBstWouldAddEndPuncttrue
\mciteSetBstMidEndSepPunct{\mcitedefaultmidpunct}
{\mcitedefaultendpunct}{\mcitedefaultseppunct}\relax
\EndOfBibitem
\bibitem[Guo \emph{et~al.}(2017)Guo, Zhou, Bai, and Zhao]{Guo2017}
Y.~Guo, S.~Zhou, Y.~Bai and J.~Zhao, \emph{ACS Appl. Mater. Interfaces}, 2017,
  \textbf{9}, 12013--12020\relax
\mciteBstWouldAddEndPuncttrue
\mciteSetBstMidEndSepPunct{\mcitedefaultmidpunct}
{\mcitedefaultendpunct}{\mcitedefaultseppunct}\relax
\EndOfBibitem
\bibitem[Hlushchenko \emph{et~al.}(2023)Hlushchenko, Siudzinska, Cybinska,
  Guzik, Bachmatiuk, and Kudrawiec]{Hlushchenko2023}
D.~Hlushchenko, A.~Siudzinska, J.~Cybinska, M.~Guzik, A.~Bachmatiuk and
  R.~Kudrawiec, \emph{Sci. Rep.}, 2023, \textbf{13}, 19114\relax
\mciteBstWouldAddEndPuncttrue
\mciteSetBstMidEndSepPunct{\mcitedefaultmidpunct}
{\mcitedefaultendpunct}{\mcitedefaultseppunct}\relax
\EndOfBibitem
\bibitem[{Komsa, Hannu-Pekka and Kotakoski, Jani and Kurasch, Simon and
  Lehtinen, Ossi and Kaiser, Ute and Krasheninnikov, Arkady
  V}(2012)]{Komsa2012}
{Komsa, Hannu-Pekka and Kotakoski, Jani and Kurasch, Simon and Lehtinen, Ossi
  and Kaiser, Ute and Krasheninnikov, Arkady V}, \emph{Phys. Rev. Lett.}, 2012,
  \textbf{109}, 035503\relax
\mciteBstWouldAddEndPuncttrue
\mciteSetBstMidEndSepPunct{\mcitedefaultmidpunct}
{\mcitedefaultendpunct}{\mcitedefaultseppunct}\relax
\EndOfBibitem
\bibitem[Zhou \emph{et~al.}(2013)Zhou, Zou, Najmaei, Liu, Shi, Kong, Lou,
  Ajayan, Yakobson, and Idrobo]{Zhou2013}
W.~Zhou, X.~Zou, S.~Najmaei, Z.~Liu, Y.~Shi, J.~Kong, J.~Lou, P.~M. Ajayan,
  B.~I. Yakobson and J.-C. Idrobo, \emph{Nano Lett.}, 2013, \textbf{13},
  2615--2622\relax
\mciteBstWouldAddEndPuncttrue
\mciteSetBstMidEndSepPunct{\mcitedefaultmidpunct}
{\mcitedefaultendpunct}{\mcitedefaultseppunct}\relax
\EndOfBibitem
\bibitem[{Hong, Jinhua and Hu, Zhixin and Probert, Matt and Li, Kun and Lv,
  Danhui and Yang, Xinan and Gu, Lin and Mao, Nannan and Feng, Qingliang and
  Xie, Liming and others}(2015)]{Hong2015}
{Hong, Jinhua and Hu, Zhixin and Probert, Matt and Li, Kun and Lv, Danhui and
  Yang, Xinan and Gu, Lin and Mao, Nannan and Feng, Qingliang and Xie, Liming
  and others}, \emph{Nat. Commun.}, 2015, \textbf{6}, 6293\relax
\mciteBstWouldAddEndPuncttrue
\mciteSetBstMidEndSepPunct{\mcitedefaultmidpunct}
{\mcitedefaultendpunct}{\mcitedefaultseppunct}\relax
\EndOfBibitem
\bibitem[Banhart \emph{et~al.}(2011)Banhart, Kotakoski, and
  Krasheninnikov]{F.Banhart2011}
F.~Banhart, J.~Kotakoski and A.~V. Krasheninnikov, \emph{ACS Nano}, 2011,
  \textbf{5}, 26--41\relax
\mciteBstWouldAddEndPuncttrue
\mciteSetBstMidEndSepPunct{\mcitedefaultmidpunct}
{\mcitedefaultendpunct}{\mcitedefaultseppunct}\relax
\EndOfBibitem
\bibitem[Hashimoto \emph{et~al.}(2004)Hashimoto, Suenaga, Gloter, Urita, and
  Iijima]{A.Hashimoto2004}
A.~Hashimoto, K.~Suenaga, A.~Gloter, K.~Urita and S.~Iijima, \emph{Nature},
  2004, \textbf{430}, 870--873\relax
\mciteBstWouldAddEndPuncttrue
\mciteSetBstMidEndSepPunct{\mcitedefaultmidpunct}
{\mcitedefaultendpunct}{\mcitedefaultseppunct}\relax
\EndOfBibitem
\bibitem[{{\"O}z{\c{c}}elik, V Ongun and Gurel, H Hakan and Ciraci,
  Salim}(2013)]{Oezcelik2013}
{{\"O}z{\c{c}}elik, V Ongun and Gurel, H Hakan and Ciraci, Salim}, \emph{Phys.
  Rev. B}, 2013, \textbf{88}, 045440\relax
\mciteBstWouldAddEndPuncttrue
\mciteSetBstMidEndSepPunct{\mcitedefaultmidpunct}
{\mcitedefaultendpunct}{\mcitedefaultseppunct}\relax
\EndOfBibitem
\bibitem[Guo \emph{et~al.}(2017)Guo, Zhou, Bai, and Zhao]{Guo2017a}
Y.~Guo, S.~Zhou, Y.~Bai and J.~Zhao, \emph{J. Chem. Phys.}, 2017, \textbf{147},
  104709\relax
\mciteBstWouldAddEndPuncttrue
\mciteSetBstMidEndSepPunct{\mcitedefaultmidpunct}
{\mcitedefaultendpunct}{\mcitedefaultseppunct}\relax
\EndOfBibitem
\bibitem[Gomes \emph{et~al.}(2016)Gomes, Carvalho, and Castro~Neto]{Gomes2016}
L.~C. Gomes, A.~Carvalho and A.~H. Castro~Neto, \emph{Phys. Rev. B}, 2016,
  \textbf{94}, 054103\relax
\mciteBstWouldAddEndPuncttrue
\mciteSetBstMidEndSepPunct{\mcitedefaultmidpunct}
{\mcitedefaultendpunct}{\mcitedefaultseppunct}\relax
\EndOfBibitem
\bibitem[Lee \emph{et~al.}(2019)Lee, Luo, Cho, Kanatzidis, and Chung]{Lee2019}
Y.~K. Lee, Z.~Luo, S.~P. Cho, M.~G. Kanatzidis and I.~Chung, \emph{Joule},
  2019, \textbf{3}, 719--731\relax
\mciteBstWouldAddEndPuncttrue
\mciteSetBstMidEndSepPunct{\mcitedefaultmidpunct}
{\mcitedefaultendpunct}{\mcitedefaultseppunct}\relax
\EndOfBibitem
\bibitem[Sutter \emph{et~al.}(2019)Sutter, Zhang, Sun, and Sutter]{Sutter2019}
E.~Sutter, B.~Zhang, M.~Sun and P.~Sutter, \emph{ACS Nano}, 2019, \textbf{13},
  9352--9362\relax
\mciteBstWouldAddEndPuncttrue
\mciteSetBstMidEndSepPunct{\mcitedefaultmidpunct}
{\mcitedefaultendpunct}{\mcitedefaultseppunct}\relax
\EndOfBibitem
\bibitem[Higashitarumizu \emph{et~al.}(2018)Higashitarumizu, Kawamoto,
  Nakamura, Shimamura, Ohashi, Ueno, and Nagashio]{Higashitarumizu2018}
N.~Higashitarumizu, H.~Kawamoto, M.~Nakamura, K.~Shimamura, N.~Ohashi, K.~Ueno
  and K.~Nagashio, \emph{Nanoscale}, 2018, \textbf{10}, 22474--22483\relax
\mciteBstWouldAddEndPuncttrue
\mciteSetBstMidEndSepPunct{\mcitedefaultmidpunct}
{\mcitedefaultendpunct}{\mcitedefaultseppunct}\relax
\EndOfBibitem
\bibitem[Li \emph{et~al.}(2016)Li, He, Heremans, and Zhao]{Li2016}
Y.~Li, B.~He, J.~P. Heremans and J.-C. Zhao, \emph{J. Alloys Compd.}, 2016,
  \textbf{669}, 224--231\relax
\mciteBstWouldAddEndPuncttrue
\mciteSetBstMidEndSepPunct{\mcitedefaultmidpunct}
{\mcitedefaultendpunct}{\mcitedefaultseppunct}\relax
\EndOfBibitem
\bibitem[Grønborg \emph{et~al.}(2019)Grønborg, Thorarinsdottir, Kyhl,
  Rodriguez-Fernández, Sanders, Bianchi, Hofmann, Miwa, Ulstrup, and
  Lauritsen]{Groenborg2019}
S.~S. Grønborg, K.~Thorarinsdottir, L.~Kyhl, J.~Rodriguez-Fernández, C.~E.
  Sanders, M.~Bianchi, P.~Hofmann, J.~A. Miwa, S.~Ulstrup and J.~V. Lauritsen,
  \emph{2D Mater.}, 2019, \textbf{6}, 045013\relax
\mciteBstWouldAddEndPuncttrue
\mciteSetBstMidEndSepPunct{\mcitedefaultmidpunct}
{\mcitedefaultendpunct}{\mcitedefaultseppunct}\relax
\EndOfBibitem
\bibitem[Xie \emph{et~al.}(2022)Xie, Li, Cheng, Haidry, Tao, Xu, Xu, and
  Ou]{xie2022recent}
H.~Xie, Z.~Li, L.~Cheng, A.~A. Haidry, J.~Tao, Y.~Xu, K.~Xu and J.~Z. Ou,
  \emph{Iscience}, 2022, \textbf{25}, 103598\relax
\mciteBstWouldAddEndPuncttrue
\mciteSetBstMidEndSepPunct{\mcitedefaultmidpunct}
{\mcitedefaultendpunct}{\mcitedefaultseppunct}\relax
\EndOfBibitem
\bibitem[Labed \emph{et~al.}(2025)Labed, Jeon, Park, Pearton, and {Seung
  Rim}]{Labed2025}
M.~Labed, H.~J. Jeon, J.~H. Park, S.~Pearton and Y.~{Seung Rim}, \emph{Mater.
  Today}, 2025, \textbf{83}, 513--537\relax
\mciteBstWouldAddEndPuncttrue
\mciteSetBstMidEndSepPunct{\mcitedefaultmidpunct}
{\mcitedefaultendpunct}{\mcitedefaultseppunct}\relax
\EndOfBibitem
\bibitem[Seo \emph{et~al.}(2017)Seo, Lee, Yoon, Park, Hwang, Kim, Yu, and
  Cho]{Seo2017}
Y.~Seo, T.~I. Lee, C.~M. Yoon, B.-E. Park, W.~S. Hwang, H.~Kim, H.-Y. Yu and
  B.~J. Cho, \emph{IEEE Trans. Electron Devices}, 2017, \textbf{64},
  3303--3307\relax
\mciteBstWouldAddEndPuncttrue
\mciteSetBstMidEndSepPunct{\mcitedefaultmidpunct}
{\mcitedefaultendpunct}{\mcitedefaultseppunct}\relax
\EndOfBibitem
\bibitem[Wei \emph{et~al.}(2024)Wei, Liu, Lan, Yang, Huang, Wang, and
  Chen]{Wei2024}
C.~Wei, J.~Liu, X.~Lan, C.~Yang, S.~Huang, X.~Wang and D.~Chen, \emph{Vacuum},
  2024, \textbf{225}, 113233\relax
\mciteBstWouldAddEndPuncttrue
\mciteSetBstMidEndSepPunct{\mcitedefaultmidpunct}
{\mcitedefaultendpunct}{\mcitedefaultseppunct}\relax
\EndOfBibitem
\bibitem[Shibayama \emph{et~al.}(2015)Shibayama, Yoshida, Kato, Sakashita,
  Takeuchi, Taoka, Nakatsuka, and Zaima]{Shibayama2015}
S.~Shibayama, T.~Yoshida, K.~Kato, M.~Sakashita, W.~Takeuchi, N.~Taoka,
  O.~Nakatsuka and S.~Zaima, \emph{Appl. Phys. Lett.}, 2015, \textbf{106},
  062107\relax
\mciteBstWouldAddEndPuncttrue
\mciteSetBstMidEndSepPunct{\mcitedefaultmidpunct}
{\mcitedefaultendpunct}{\mcitedefaultseppunct}\relax
\EndOfBibitem
\bibitem[Miller \emph{et~al.}(2020)Miller, Chesaux, Deligiannis, Mascher, and
  Bradley]{smiller2020low}
J.~Miller, M.~Chesaux, D.~Deligiannis, P.~Mascher and J.~Bradley, \emph{Thin
  Solid Films}, 2020, \textbf{709}, 138165\relax
\mciteBstWouldAddEndPuncttrue
\mciteSetBstMidEndSepPunct{\mcitedefaultmidpunct}
{\mcitedefaultendpunct}{\mcitedefaultseppunct}\relax
\EndOfBibitem
\bibitem[Mizoguchi \emph{et~al.}(2022)Mizoguchi, Ishiyama, Moto, Imajo,
  Suemasu, and Toko]{mizoguchi2022solid}
T.~Mizoguchi, T.~Ishiyama, K.~Moto, T.~Imajo, T.~Suemasu and K.~Toko,
  \emph{PHYS STATUS SOLIDI-R}, 2022, \textbf{16}, 2100509\relax
\mciteBstWouldAddEndPuncttrue
\mciteSetBstMidEndSepPunct{\mcitedefaultmidpunct}
{\mcitedefaultendpunct}{\mcitedefaultseppunct}\relax
\EndOfBibitem
\bibitem[Liu \emph{et~al.}(2025)Liu, Wei, Yang, Lan, Huang, Zhang, and
  Wang]{liu2025high}
J.~Liu, C.~Wei, C.~Yang, X.~Lan, S.~Huang, F.~Zhang and X.~Wang, \emph{Solid
  State Commun.}, 2025,  115856\relax
\mciteBstWouldAddEndPuncttrue
\mciteSetBstMidEndSepPunct{\mcitedefaultmidpunct}
{\mcitedefaultendpunct}{\mcitedefaultseppunct}\relax
\EndOfBibitem
\bibitem[Guo \emph{et~al.}(2019)Guo, Ma, Mao, Ju, Bai, Zhao, and Zeng]{Guo2019}
Y.~Guo, L.~Ma, K.~Mao, M.~Ju, Y.~Bai, J.~Zhao and X.~C. Zeng, \emph{Nanoscale
  Horiz.}, 2019, \textbf{4}, 592--600\relax
\mciteBstWouldAddEndPuncttrue
\mciteSetBstMidEndSepPunct{\mcitedefaultmidpunct}
{\mcitedefaultendpunct}{\mcitedefaultseppunct}\relax
\EndOfBibitem
\bibitem[Zhang \emph{et~al.}(2021)Zhang, Xu, Yao, Jannat, Ren, Field, Wen,
  Zhou, Zavabeti, and Ou]{Zhang2021}
B.~Y. Zhang, K.~Xu, Q.~Yao, A.~Jannat, G.~Ren, M.~R. Field, X.~Wen, C.~Zhou,
  A.~Zavabeti and J.~Z. Ou, \emph{Nat. Mater.}, 2021, \textbf{20},
  1073--1078\relax
\mciteBstWouldAddEndPuncttrue
\mciteSetBstMidEndSepPunct{\mcitedefaultmidpunct}
{\mcitedefaultendpunct}{\mcitedefaultseppunct}\relax
\EndOfBibitem
\bibitem[Sozen \emph{et~al.}(2021)Sozen, Yagmurcukardes, and Sahin]{Sozen2021}
Y.~Sozen, M.~Yagmurcukardes and H.~Sahin, \emph{Phys. Chem. Chem. Phys.}, 2021,
  \textbf{23}, 21307--21315\relax
\mciteBstWouldAddEndPuncttrue
\mciteSetBstMidEndSepPunct{\mcitedefaultmidpunct}
{\mcitedefaultendpunct}{\mcitedefaultseppunct}\relax
\EndOfBibitem
\bibitem[Wan \emph{et~al.}(2024)Wan, Peng, Ge, Fu, and Liu]{Wan2024}
W.~Wan, Y.~Peng, Y.~Ge, B.~Fu and Y.~Liu, \emph{Physica E}, 2024, \textbf{162},
  115997\relax
\mciteBstWouldAddEndPuncttrue
\mciteSetBstMidEndSepPunct{\mcitedefaultmidpunct}
{\mcitedefaultendpunct}{\mcitedefaultseppunct}\relax
\EndOfBibitem
\bibitem[Riaz \emph{et~al.}(2023)Riaz, Gul, Khan, Ahmad, and Ilyas]{Riaz2023}
S.~Riaz, M.~Gul, F.~Khan, I.~Ahmad and M.~Ilyas, \emph{Appl. Phys. A}, 2023,
  \textbf{129}, 589\relax
\mciteBstWouldAddEndPuncttrue
\mciteSetBstMidEndSepPunct{\mcitedefaultmidpunct}
{\mcitedefaultendpunct}{\mcitedefaultseppunct}\relax
\EndOfBibitem
\bibitem[Li \emph{et~al.}(2023)Li, Yuan, Li, Liu, Shen, Jiang, Song, and
  Xia]{li2023two}
X.~Li, P.~Yuan, L.~Li, T.~Liu, C.~Shen, Y.~Jiang, X.~Song and C.~Xia,
  \emph{Front. Phys.}, 2023, \textbf{18}, 13305\relax
\mciteBstWouldAddEndPuncttrue
\mciteSetBstMidEndSepPunct{\mcitedefaultmidpunct}
{\mcitedefaultendpunct}{\mcitedefaultseppunct}\relax
\EndOfBibitem
\bibitem[Kresse and Furthm\"uller(1996)]{b2}
G.~Kresse and J.~Furthm\"uller, \emph{Phys. Rev. B}, 1996, \textbf{54},
  11169--11186\relax
\mciteBstWouldAddEndPuncttrue
\mciteSetBstMidEndSepPunct{\mcitedefaultmidpunct}
{\mcitedefaultendpunct}{\mcitedefaultseppunct}\relax
\EndOfBibitem
\bibitem[Kresse and Joubert(1999)]{Kresse1999}
G.~Kresse and D.~Joubert, \emph{Phys. Rev. B}, 1999, \textbf{59}, 1758\relax
\mciteBstWouldAddEndPuncttrue
\mciteSetBstMidEndSepPunct{\mcitedefaultmidpunct}
{\mcitedefaultendpunct}{\mcitedefaultseppunct}\relax
\EndOfBibitem
\bibitem[Perdew \emph{et~al.}(1996)Perdew, Burke, and Ernzerhof]{b4}
J.~P. Perdew, K.~Burke and M.~Ernzerhof, \emph{Phys. Rev. Lett.}, 1996,
  \textbf{77}, 3865\relax
\mciteBstWouldAddEndPuncttrue
\mciteSetBstMidEndSepPunct{\mcitedefaultmidpunct}
{\mcitedefaultendpunct}{\mcitedefaultseppunct}\relax
\EndOfBibitem
\bibitem[Monkhorst and Pack(1976)]{b5}
H.~J. Monkhorst and J.~D. Pack, \emph{Phys. Rev. B}, 1976, \textbf{13},
  5188\relax
\mciteBstWouldAddEndPuncttrue
\mciteSetBstMidEndSepPunct{\mcitedefaultmidpunct}
{\mcitedefaultendpunct}{\mcitedefaultseppunct}\relax
\EndOfBibitem
\bibitem[Grimme(2006)]{Chem.2006}
S.~Grimme, \emph{J. Comput. Chem.}, 2006, \textbf{27}, 1787--1799\relax
\mciteBstWouldAddEndPuncttrue
\mciteSetBstMidEndSepPunct{\mcitedefaultmidpunct}
{\mcitedefaultendpunct}{\mcitedefaultseppunct}\relax
\EndOfBibitem
\bibitem[Henkelman \emph{et~al.}(2000)Henkelman, Uberuaga, and
  J{\'o}nsson]{Henkelman2000}
G.~Henkelman, B.~P. Uberuaga and H.~J{\'o}nsson, \emph{J. Chem. Phys.}, 2000,
  \textbf{113}, 9901--9904\relax
\mciteBstWouldAddEndPuncttrue
\mciteSetBstMidEndSepPunct{\mcitedefaultmidpunct}
{\mcitedefaultendpunct}{\mcitedefaultseppunct}\relax
\EndOfBibitem
\bibitem[Naclerio \emph{et~al.}(2020)Naclerio, Zakharov, Kumar, Rogers, Pint,
  Shrivastava, and Kidambi]{Naclerio2020}
A.~E. Naclerio, D.~N. Zakharov, J.~Kumar, B.~R. Rogers, C.~L. Pint,
  M.~Shrivastava and P.~R. Kidambi, \emph{ACS Appl. Mater. Interfaces},
  2020\relax
\mciteBstWouldAddEndPuncttrue
\mciteSetBstMidEndSepPunct{\mcitedefaultmidpunct}
{\mcitedefaultendpunct}{\mcitedefaultseppunct}\relax
\EndOfBibitem
\bibitem[Dong \emph{et~al.}(2021)Dong, Zhou, Xin, Yang, Zhang, Liu, Zhang,
  Yang, and Liu]{dong2021investigations}
L.~Dong, S.~Zhou, B.~Xin, C.~Yang, J.~Zhang, H.~Liu, L.~Zhang, C.~Yang and
  W.~Liu, \emph{Appl. Surf. Sci.}, 2021, \textbf{537}, 147883\relax
\mciteBstWouldAddEndPuncttrue
\mciteSetBstMidEndSepPunct{\mcitedefaultmidpunct}
{\mcitedefaultendpunct}{\mcitedefaultseppunct}\relax
\EndOfBibitem
\bibitem[Guo \emph{et~al.}(2022)Guo, Zhao, Zhou, and Zhao]{guo2022oxidation}
Y.~Guo, Y.~Zhao, S.~Zhou and J.~Zhao, \emph{NANOSCALE}, 2022, \textbf{14},
  11452--11460\relax
\mciteBstWouldAddEndPuncttrue
\mciteSetBstMidEndSepPunct{\mcitedefaultmidpunct}
{\mcitedefaultendpunct}{\mcitedefaultseppunct}\relax
\EndOfBibitem
\bibitem[Rawat \emph{et~al.}(2024)Rawat, Patra, Pandey, and
  Karna]{rawat2024first}
A.~Rawat, L.~Patra, R.~Pandey and S.~P. Karna, \emph{NANOSCALE}, 2024,
  \textbf{16}, 7437--7442\relax
\mciteBstWouldAddEndPuncttrue
\mciteSetBstMidEndSepPunct{\mcitedefaultmidpunct}
{\mcitedefaultendpunct}{\mcitedefaultseppunct}\relax
\EndOfBibitem
\bibitem[Wu \emph{et~al.}(2019)Wu, Xu, Lin, Wang, and Zeng]{wu2019two}
Q.~Wu, W.~W. Xu, D.~Lin, J.~Wang and X.~C. Zeng, \emph{J. Phys. Chem. Lett.},
  2019, \textbf{10}, 3773--3778\relax
\mciteBstWouldAddEndPuncttrue
\mciteSetBstMidEndSepPunct{\mcitedefaultmidpunct}
{\mcitedefaultendpunct}{\mcitedefaultseppunct}\relax
\EndOfBibitem
\bibitem[Guo \emph{et~al.}(2019)Guo, Wu, Li, Lu, Mao, Bai, Zhao, Wang, and
  Zeng]{Guo2019a}
Y.~Guo, Q.~Wu, Y.~Li, N.~Lu, K.~Mao, Y.~Bai, J.~Zhao, J.~Wang and X.~C. Zeng,
  \emph{Nanoscale Horiz.}, 2019, \textbf{4}, 223--230\relax
\mciteBstWouldAddEndPuncttrue
\mciteSetBstMidEndSepPunct{\mcitedefaultmidpunct}
{\mcitedefaultendpunct}{\mcitedefaultseppunct}\relax
\EndOfBibitem
\bibitem[Speight(2016)]{Speight2016}
J.~Speight, \emph{Lange's Handbook of Chemistry, Seventeenth Edition},
  McGraw-Hill Education, New York, N.Y, 2016\relax
\mciteBstWouldAddEndPuncttrue
\mciteSetBstMidEndSepPunct{\mcitedefaultmidpunct}
{\mcitedefaultendpunct}{\mcitedefaultseppunct}\relax
\EndOfBibitem
\bibitem[Shukla and Gaur(2020)]{shukla2020dft}
A.~Shukla and N.~Gaur, \emph{Chem. Phys. Lett.}, 2020, \textbf{754},
  137717\relax
\mciteBstWouldAddEndPuncttrue
\mciteSetBstMidEndSepPunct{\mcitedefaultmidpunct}
{\mcitedefaultendpunct}{\mcitedefaultseppunct}\relax
\EndOfBibitem
\bibitem[Levine \emph{et~al.}(2019)Levine, Vera, Kulbak, Ceratti, Rehermann,
  Márquez, Levcenko, Unold, Hodes, Balberg, Cahen, and Dittrich]{Levine2019}
I.~Levine, O.~G. Vera, M.~Kulbak, D.-R. Ceratti, C.~Rehermann, J.~A. Márquez,
  S.~Levcenko, T.~Unold, G.~Hodes, I.~Balberg, D.~Cahen and T.~Dittrich,
  \emph{ACS Energy Lett.}, 2019, \textbf{4}, 1150--1157\relax
\mciteBstWouldAddEndPuncttrue
\mciteSetBstMidEndSepPunct{\mcitedefaultmidpunct}
{\mcitedefaultendpunct}{\mcitedefaultseppunct}\relax
\EndOfBibitem
\bibitem[Ziletti \emph{et~al.}(2015)Ziletti, Carvalho, Campbell, Coker, and
  Castro~Neto]{Ziletti2015}
A.~Ziletti, A.~Carvalho, D.~K. Campbell, D.~F. Coker and A.~H. Castro~Neto,
  \emph{Phys. Rev. Lett.}, 2015, \textbf{114}, 046801\relax
\mciteBstWouldAddEndPuncttrue
\mciteSetBstMidEndSepPunct{\mcitedefaultmidpunct}
{\mcitedefaultendpunct}{\mcitedefaultseppunct}\relax
\EndOfBibitem
\bibitem[KC \emph{et~al.}(2015)KC, Longo, Wallace, and Cho]{KC2015}
S.~KC, R.~C. Longo, R.~M. Wallace and K.~Cho, \emph{J. Appl. Phys.}, 2015,
  \textbf{117}, 135301\relax
\mciteBstWouldAddEndPuncttrue
\mciteSetBstMidEndSepPunct{\mcitedefaultmidpunct}
{\mcitedefaultendpunct}{\mcitedefaultseppunct}\relax
\EndOfBibitem
\bibitem[Jiang \emph{et~al.}(2018)Jiang, Zhang, Ao, Huang, He, and
  Wu]{Jiang2018}
Q.~Jiang, J.~Zhang, Z.~Ao, H.~Huang, H.~He and Y.~Wu, \emph{Front. Chem.},
  2018, \textbf{6}, 187\relax
\mciteBstWouldAddEndPuncttrue
\mciteSetBstMidEndSepPunct{\mcitedefaultmidpunct}
{\mcitedefaultendpunct}{\mcitedefaultseppunct}\relax
\EndOfBibitem
\bibitem[Huang \emph{et~al.}(2021)Huang, Song, Zhang, Zhao, Hou, Hoang, and
  Chen]{huang2021defects}
W.~Huang, M.~Song, Y.~Zhang, Y.~Zhao, H.~Hou, L.~H. Hoang and X.~Chen,
  \emph{Opt. Mater.}, 2021, \textbf{119}, 111372\relax
\mciteBstWouldAddEndPuncttrue
\mciteSetBstMidEndSepPunct{\mcitedefaultmidpunct}
{\mcitedefaultendpunct}{\mcitedefaultseppunct}\relax
\EndOfBibitem
\bibitem[Intayot and Jungsuttiwong(2025)]{intayot2025unveiling}
R.~Intayot and S.~Jungsuttiwong, \emph{J. Environ. Chem. Eng.}, 2025,
  116569\relax
\mciteBstWouldAddEndPuncttrue
\mciteSetBstMidEndSepPunct{\mcitedefaultmidpunct}
{\mcitedefaultendpunct}{\mcitedefaultseppunct}\relax
\EndOfBibitem
\bibitem[He \emph{et~al.}(2018)He, Lu, and Ma]{he2018interaction}
B.~He, Z.~Lu and D.~Ma, \emph{Vacuum}, 2018, \textbf{153}, 53--61\relax
\mciteBstWouldAddEndPuncttrue
\mciteSetBstMidEndSepPunct{\mcitedefaultmidpunct}
{\mcitedefaultendpunct}{\mcitedefaultseppunct}\relax
\EndOfBibitem
\bibitem[Poldorn \emph{et~al.}(2020)Poldorn, Wongnongwa, Namuangruk, Kungwan,
  Golovko, Inceesungvorn, and Jungsuttiwong]{poldorn2020theoretical}
P.~Poldorn, Y.~Wongnongwa, S.~Namuangruk, N.~Kungwan, V.~B. Golovko,
  B.~Inceesungvorn and S.~Jungsuttiwong, \emph{APPL CATAL A-GEN}, 2020,
  \textbf{595}, 117505\relax
\mciteBstWouldAddEndPuncttrue
\mciteSetBstMidEndSepPunct{\mcitedefaultmidpunct}
{\mcitedefaultendpunct}{\mcitedefaultseppunct}\relax
\EndOfBibitem
\bibitem[Da~Silva \emph{et~al.}(2012)Da~Silva, Rolim, Soares, Baumvol, Krug,
  Miotti, Freire, Da~Costa, and Radtke]{DaSilva2012}
S.~Da~Silva, G.~K. Rolim, G.~V. Soares, I.~J.~R. Baumvol, C.~Krug, L.~Miotti,
  F.~Freire, M.~Da~Costa and C.~Radtke, \emph{Appl. Phys. Lett.}, 2012,
  \textbf{100}, 191907\relax
\mciteBstWouldAddEndPuncttrue
\mciteSetBstMidEndSepPunct{\mcitedefaultmidpunct}
{\mcitedefaultendpunct}{\mcitedefaultseppunct}\relax
\EndOfBibitem
\bibitem[{Al Ghaithi} \emph{et~al.}(2025){Al Ghaithi}, Taha, Ansari, Rajput,
  Mohammad, and Aldosari]{AlGhaithi2025}
A.~O. {Al Ghaithi}, I.~Taha, S.~M. Ansari, N.~Rajput, B.~Mohammad and H.~M.
  Aldosari, \emph{Vacuum}, 2025, \textbf{231}, 113791\relax
\mciteBstWouldAddEndPuncttrue
\mciteSetBstMidEndSepPunct{\mcitedefaultmidpunct}
{\mcitedefaultendpunct}{\mcitedefaultseppunct}\relax
\EndOfBibitem
\end{mcitethebibliography}



\end{document}